\documentclass{osa-article}

\journal{osac}

\usepackage{siunitx}
\usepackage[position=top]{subfig}
\usepackage{graphicx,wrapfig}

\articletype{Research Article}

\begin{document}

\title{Numerical analysis of $\text{LG}_{3,3}$ second harmonic generation in comparison to the $\text{LG}_{0,0}$ case}

\author{Joscha Heinze,\authormark{1,*} Henning Vahlbruch,\authormark{1} and Benno Willke\authormark{1}}

\address{\authormark{1}Max-Planck-Institut für Gravitationsphysik (Albert-Einstein-Institut) and Leibniz Universität Hannover, 30167 Hannover, Germany}

\email{\authormark{*}joscha.heinze@aei.mpg.de} 



\begin{abstract}
For coating Brownian thermal noise reduction in future gravitational wave detectors, it is proposed to use light in the helical Laguerre-Gaussian $\text{LG}_{3,3}$ mode instead of the currently used $\text{LG}_{0,0}$ mode. However, the simultaneous reduction of quantum noise would then require the efficient generation of squeezed vacuum states in the $\text{LG}_{3,3}$ mode. Current squeezed light generation techniques employ continuous-wave second harmonic generation (SHG). Here, we simulate the SHG for both modes numerically to derive first insights into the transferability of standard squeezed light generation techniques to the $\text{LG}_{3,3}$ mode. In the first part of this paper, we therefore theoretically discuss SHG in the case of a single undepleted pump mode, which, in general, excites a superposition of harmonic modes. Based on the differential equation for the harmonic field, we derive individual phase matching conditions and hence conversion efficiencies for the excited harmonic modes. In the second part, we analyse the numerical simulations of the $\text{LG}_{0,0}$ and $\text{LG}_{3,3}$ SHG in a single-pass, double-pass and cavity-enhanced configuration under the influence of the focusing, the different pump intensity distributions and the individual phase matching conditions. Our results predict that the $\text{LG}_{3,3}$ mode requires about 14 times the pump power of the $\text{LG}_{0,0}$ mode to achieve the same SHG conversion efficiency in an ideal, realistic cavity design and mainly generates the harmonic $\text{LG}_{6,6}$ mode.
\end{abstract}

\section{Introduction}
One main focus of the gravitational wave community is the steady improvement of the gravitational wave detectors. Besides e.g.\ quantum and seismic noise, which are relevant in different frequency ranges, coating Brownian thermal noise fundamentally limits the sensitivity of second generation detectors such as Advanced LIGO \cite{Intro:aLIGO} and Advanced Virgo \cite{Intro:AdV} around \SI{100}{Hz}. One proposed method to reduce its coupling to the sensitivity of future detectors is the change from the currently used fundamental Gaussian mode to higher-order Laguerre-Gaussian (LG) modes \cite{Intro:reduceThermalNoise}. The high-purity generation of the helical $\text{LG}_{3,3}$ mode and its general compatibility with the design of gravitational wave detectors has already been demonstrated \cite{Birmingham-highpurity,prospects,highPowerHighPurity,Noack,MichelsonInterferometer_LG33}, while challenges with respect to mode degeneracy still exist \cite{10mHighFinesseCavity_LG33,degeneracyHighFinesseCavity}. On the other hand, the efficiency of the $\text{LG}_{3,3}$ mode in the quantum noise reduction via squeezed vacuum states, as currently done with the fundamental Gaussian mode \cite{AdVirgoSqueezing}, has not been examined yet. We investigate how the concept of producing squeezed vacuum states via second harmonic generation (SHG) and subsquent parametric down-conversion can be adapted to the $\text{LG}_{3,3}$ mode. Based on numerical simulations, we therefore compare the single-, double-pass (referring to one or two passes of the fundamental pump field through the nonlinear crystal) and cavity-enhanced SHG continuously pumped by the $\text{LG}_{0,0}$ and $\text{LG}_{3,3}$ mode.

In Sections \ref{sec:LGmode} to \ref{sec:changeInPowerAndPhase}, we derive the iterative equation for the numerical simulation of the single- and double-pass SHG from the differential equation of the harmonic field. Here, the general case of a single undepleted pump mode which excites a superposition of harmonic modes is considered. Without loss of generality, the derivation assumes the Laguerre-Gaussian mode basis due to the later mode comparison while similarities and differences to the Hermite-Gaussian (HG) basis are indicated. Furthermore, the iterative equation shows how the pump intensity distribution, the focusing (cf.\ \cite{BoydKleinmann,lastzkaGouy}) and the Gouy phase (cf.\ \cite{SHGmodeDiscrimination,SHGindividualPhaseMatching}) affect the conversion efficiency and how the Gouy phase leads to different phase matching conditions of the excited harmonic modes. In Sections \ref{sec:LG00} to \ref{sec:doublePass}, the numerical simulations of the single-pass SHG generally illustrate these correlations and provide a detailed focus on the individual phase matching conditions. These conditions are especially important for the $\text{LG}_{3,3}$ SHG and are reflected in the phase relations of the excited harmonic modes to the crystal polarisation along the crystal. Together with the subsequent double-pass simulations, our results allow for predictions regarding both the harmonic output fields and the ratio of the pump-mode dependent effective nonlinearities of the nonlinear medium. We use the latter to quantify the SHG comparison. Finally, we take the double-pass results as an input for an existing simulation tool for the cavity-enhanced SHG to compute the conversion efficiencies of both the $\text{LG}_{0,0}$ and $\text{LG}_{3,3}$ mode under typical experimental conditions prevailing in gravitational wave detectors \cite{CoherentControlBroadbandVacuumSqueezing}. 

\section{The helical Laguerre-Gaussian modes}\label{sec:LGmode}
The helical Laguerre-Gaussian modes can be derived as a solution to the paraxial Helmholtz equation (PHE). Their complete normalised form in cylindrical coordinates ($\textbf{r}=(r,\phi,z)$, $z$ defines the propagation axis) reads \cite{Siegmann-Laser,Birmingham-highpurity}
\begin{align}
\begin{split}
\text{LG}_{p,l}(\textbf{r},t)&=\frac{1}{w(z)}\sqrt{\frac{2p!}{\pi(p+\left|l\right|)!}}\left(\frac{\sqrt{2}r}{w(z)}\right)^{\left|l\right|}L_p^{\left|l\right|}\left(\frac{2r^2}{w^2(z)}\right)\\
&\quad\times e^{-ik\frac{r^2}{2q(z)}+il\phi}e^{i(2p+\left|l\right|+1)\Psi(z)}e^{i(kz-\omega t+\beta)}\\
&=A_{p,l}(\textbf{r})\times e^{i(kz-\omega t+\beta)}\\
&=T_{p,l}(\textbf{r})\times\Phi_{p,l}(z)\times e^{i(kz-\omega t+\beta)}\\
\end{split}
\label{eqn:LGmode}
\end{align}
with
\begin{align}
\begin{split}
T_{p,l}(\textbf{r})&:=\frac{1}{w(z)}\sqrt{\frac{2p!}{\pi(p+\left|l\right|)!}}\left(\frac{\sqrt{2}r}{w(z)}\right)^{\left|l\right|}L_p^{\left|l\right|}\left(\frac{2r^2}{w^2(z)}\right)\times e^{-ik\frac{r^2}{2q(z)}+il\phi}\\
\Phi_{p,l}(z)&:=e^{i(2p+\left|l\right|+1)\Psi(z)}
\end{split}
\end{align}
where $w(z)$ is the beam radius, $L_p^{\left|l\right|}$ is the generalised Laguerre polynomial, $q(z)$ is the complex beam parameter, $\Psi(z):=~\text{atan}(z/z_R)$ with the Rayleigh range $z_R=~n\pi w_0^2/\lambda$ ($n$: refractive index, $w_0$: waist size, $\lambda$: wavelength), $k=2\pi n/\lambda$ is the wavenumber and $\omega$ is the angular frequency. Without loss of generality, the additional phase term can be set to $\beta=0$. Furthermore, $p\geq 0$ is the radial and $l\in\mathbb{Z}$ is the azimuthal mode index. The complete expression for $\text{LG}_{p,l}$ can be written as the product of the amplitude distribution $A_{p,l}$, which is the solution to the PHE, and the phase term $\exp[i(kz-\omega t)]$, which is separated from $A_{p,l}$ before the PHE is solved. The amplitude distribution $A_{p,l}$ can further be separated into the normalised transverse amplitude distribution $T_{p,l}(\textbf{r})$, which includes the normalisation factors and defines the mode, and the Gouy phase term $\Phi_{p,l}(z)$, which is a global (universal in the transverse plane) phase of $A_{p,l}$ that depends on the mode order $g_{p,l}=~2p+\left|l\right|$ and the $z$-position. This last separation will help to understand how the fundamental and harmonic field interact in second harmonic generation.

In the HG basis, the expression of an arbitrary mode $\text{HG}_{m,n}$ can be separated in the same manner with a corresponding $T_{m,n}$ term and $\Phi_{m,n}(z)=~\!\exp[i(m+~\!n+~\!1)\Psi(z)]$.

\section{The differential equation}\label{sec:DEQ}
In order to develop a basic understanding of how the conversion from the fundamental pump field into the (second) harmonic field works (also refer to \cite{Boyd-nonlinearOptics}), we assume Type-I phase matching, no pump-depletion and no initial harmonic field. For simplicity, the polarisation direction of the two fields is omitted and walk-off effects and absorption are neglected. Along the $z$-axis, the change in the harmonic field is then described by the following PHE where the dielectric polarisation of the nonlinear crystal, excited by the fundamental field, serves as a source term on the right-hand side \cite{Boyd-nonlinearOptics,SHGmodeDiscrimination}
\begin{equation}
\Big[\nabla_T^2-2ik_2\partial_z\Big]A_2(\textbf{r})=\underbrace{-2\frac{\omega_2^2}{c^2}d_\text{eff}A_1^2(\textbf{r})e^{i\Delta kz}}_{=:\ S(\textbf{r})}\ \text{.}
\label{DEQ:initialForm}
\end{equation}
Here, the indices $1$ and $2$ denote the fundamental pump field and harmonic field, respectively, $c$ is the speed of light in vacuum, $d_\text{eff}$ is the effective nonlinearity of the crystal medium and $\Delta k=\!~2k_1-k_2$ is the wavevector mismatch which is synonymous to a difference in the refractive indices according to $\Delta k=~4\pi\Delta n/\lambda_1$ (for SHG) with $\Delta n=~n_1-n_2$. While the PHE, in principle, only deals with the amplitude distributions $A_1$ and $A_2$ of the fundamental and harmonic field, respectively, the dependency on the phase term $\exp(ikz)$ is still included on the right-hand side as the wavevector mismatch. The term $\exp(-i\omega t)$ from Eq.\ \ref{eqn:LGmode} is not taken into account because the SHG process is considered to be time-invariant.

As a first step to describe the SHG process, the $z$-derivative in Eq.\ \ref{DEQ:initialForm} is discretised as follows
\begin{equation}
\partial_zA_2(r,\phi,z)=\frac{A_2(r,\phi,z+\Delta z)-A_2(r,\phi,z)}{\Delta z}\ \text{.}
\label{DEQ:discreteZ}
\end{equation} 
We can apply this to the homogeneous PHE (right-hand side is zero) that describes the mere propagation of the field $A_2$:
\begin{equation}
A_2(r,\phi,z+\Delta z)=A_2(r,\phi,z)-k'\Delta z\cdot\nabla_T^2A_2(r,\phi,z)	
\label{DEQ:propagation}
\end{equation}
with $k':=i/(2k_2)$, which shows that the transverse derivative $\nabla_T^2$ determines how any superposition of modes locally changes phase and shape when propagating. We can then apply Eqs.\ \ref{DEQ:discreteZ} and \ref{DEQ:propagation} to the heterogeneous PHE:
\begin{align}
\begin{split}
A_2(r,\phi,z+\Delta z)&=A_2(r,\phi,z)-k'\Delta z\nabla_T^2\cdot A_2(r,\phi,z)+k'\Delta z\cdot S(r,\phi,z+\Delta z)\\
&=A_2'(r,\phi,z+\Delta z)+C^h(r,\phi,z+\Delta z)
\end{split}
\label{DEQ:discreteHelmholtz}
\end{align}
with
\begin{align}
\begin{split}
A_2'(r,\phi,z+\Delta z):&=A_2(r,\phi,z)-k'\Delta z\nabla_T^2\cdot A_2(r,\phi,z)\\
C^h(r,\phi,z+\Delta z):&=k'\Delta z\cdot S(r,\phi,z+\Delta z)\ \text{.}
\end{split}
\end{align}
This equation works in the following way: the harmonic field $A'_2$ which is present at $z$ propagates up to $z+\Delta z$ and interferes with the harmonic field $C^h$ that is emitted from the crystal at this subsequent position (to realise the correct interference, $\Delta z$ has to be added in the arguments of $S$ and $C^h$ which becomes negligible when $\Delta z\rightarrow~dz$). Then, the resulting harmonic field $A_2$ continues to propagate. Based on this mechanism, the numerical simulations below will compute the evolution of the generated harmonic field $A_2$ by iteratively progressing through the crystal. $\Delta z$ determines the $z$-resolution which we increased until the solution was stable. Since $C^h$ is, for now, the only term associated with the crystal polarisation, it will simply be referred to as ``crystal polarisation''. $h$ indicates the harmonic frequency.

At the beginning of the crystal $z_0$, where no harmonic field has yet been generated ($A'_2=0$), Eq.\ \ref{DEQ:discreteHelmholtz} becomes
\begin{equation}
A_2(r,\phi,z_0)=C^h(r,\phi,z_0)\ \text{.}
\label{DEQ:firstStep}
\end{equation}
Importantly, the crystal polarisation imprints its mode composition on the harmonic field $A_2$ in this initial iteration step. 

\section{The crystal polarisation}\label{sec:sourceTerm}
We assume that the SHG is pumped by a single $\text{LG}_{p,l}$ mode. According to Eq.\ \ref{DEQ:initialForm} with $A_1\propto A_{p,l}$, the crystal polarisation is then proportional to the square of $A_{p,l}$, that is the product of the squared normalised transverse amplitude distribution (nTAD) $T_{p,l}$ and squared Gouy phase $\Phi_{p,l}$, and can therefore be written as (see \cite{SHG1997} for second line)
\begin{align}
\begin{split}
C^h(\textbf{r})&\propto \big(T_{p,l}(\textbf{r})\big)^2\times\Phi_S(z)\\
\big(T_{p,l}(\textbf{r})\big)^2&=\sum_{m=0}^{p}t_{2m,2l}(z)T^h_{2m,2l}(\textbf{r})\\
\Phi_S(z)&=\big(\Phi_{p,l}(z)\big)^2=e^{i2(2p+\left|l\right|+1)\Psi(z)}\ \text{.}
\end{split}
\label{DEQ:SourceTerm}
\end{align}
where $t_{2m,2l}(z)$ are real-valued coefficients and $(T_{p,l}(\textbf{r}))^2$ is not normalised. The superscript $h$ denotes the harmonic nTADs which are characterised by the harmonic frequency and a reduced waist size compared to the fundamental nTADs ($w_0\rightarrow~2^{-1/2}w_0$). Independent on the $z$-position inside the crystal, the crystal polarisation is thus in a varying superposition of the same harmonic LG modes with $l'=2l$ (conservation of orbital angular momemtum \cite{SHG1997}) and $p'=0,2,\dots,2p$. $C^h$ consists of more than one mode if $p>0$.

Combining Eqs.\ \ref{DEQ:discreteHelmholtz}, \ref{DEQ:firstStep} and \ref{DEQ:SourceTerm} leads to the following mode-resolved version of Eq.\ \ref{DEQ:discreteHelmholtz} which iteratively shows that the harmonic field and the crystal polarisation are in superpositions of the same harmonic modes throughout the crystal (This is only strictly true for $\Delta n=0$, but still a good approximation for a reasonable small $\Delta n$, see Sec.\ \ref{sec:LG00}.):
\begin{align}
\begin{split}
A_2(r,\phi,z+\Delta z)&=\sum_{m=0}^{p}\underbrace{\Big[a_{2m,l'}(z)+c_{2m,l'}(z+\Delta z)\Big]}_{=\ a_{2m,l'}(z+\Delta z)}\times A^h_{2m,l'}(r,\phi,z+\Delta z)
\end{split}
\label{DEQ:discreteHelmholtzWithModes}
\end{align}
where $a_{2m,l'}(z)$ are the complex coefficients for the harmonic field which has been generated from $z_0$ up to $z$. These coefficients do not change with the mere propagation along $\Delta z$ which is completely described by $A^h_{2m,l'}$. The complex $c_{2m,l'}(z+\Delta z)$ coefficients belong to the harmonic field emitted from the crystal polarisation at $z+\Delta z$. Since the LG modes form an orthonormal basis, the interference in each term can be analysed individually.

Eqs.\ \ref{DEQ:SourceTerm} and \ref{DEQ:discreteHelmholtzWithModes} also apply to the HG basis by considering the corresponding expressions for $(T_{m,n})^2$ and $\Phi_S$. 

\section{Phase matching}\label{sec:phaseMatching}
Phase matching refers to the evolution of the phase difference $\Delta\alpha_{2m,l'}(z)\in(-\pi,\pi]$ between $a_{2m,l'}(z)$ and $c_{2m,l'}(z+\Delta z)$ in Eq.\ \ref{DEQ:discreteHelmholtzWithModes} where a ``good'' phase matching is characterised by $\Delta\alpha_{2m,l'}(z)\approx~0$ along the crystal. The evolution of this phase difference is determined by the global phase terms in the harmonic field $A_2$ and crystal polarisation $C^h$. 

The harmonic field includes the respective nTADs and Gouy phases of the involved harmonic modes:
\begin{equation}
A_2(\textbf{r})=\sum_{m=0}^{p}a_{2m,l'}(z)T^h_{2m,l'}(\textbf{r})\Phi_{2m,l'}(z)\ \text{.}
\label{DEQ:harmonicField}
\end{equation}
The crystal polarisation consists of the same nTADs, but since it originates from squaring the pump mode, they are all weighted by the same Gouy phase (and wavevector mismatch):
\begin{align}
\begin{split}
&C^h(\textbf{r})=i\mathcal{C}_{p,l}(z)\left[\sum_{m=0}^{p}c'_{2m,l'}T^h_{2m,l'}(\textbf{r})\right]\Phi_S(z)e^{i\Delta k z}\ \text{,}\\
&\sum_{m=0}^{p}\left|c'_{2m,l'}\right|^2=1\ \text{.}
\end{split}
\label{DEQ:sourceTermComplete}
\end{align}
where the constant real-valued $c'_{2m,l'}$ coefficients quantify the fractional contributions of the harmonic modes to the crystal polarisation and $\mathcal{C}_{p,l}(z)$ is the real-valued amplitude. 

This discrepancy in the Gouy phases which is basically a discrepancy in the Gouy phase factors $\gamma_{p,l}:=~2p+\left|l\right|+1$ leads to an interesting effect. On the one hand, the nTADs (the modes) in the harmonic field and crystal polarisation are identical throughout the crystal. On the other hand, the single Gouy phase of the crystal polarisation evolves faster than each Gouy phase in the harmonic field. For each harmonic mode, an individual Gouy phase difference between the already generated harmonic field and the crystal polarisation will consequently build up during the propagation through the crystal. Hence, the initially constructive interferences in Eq.\ \ref{DEQ:discreteHelmholtzWithModes} experience an individually fast decaying dependent on $\Delta\gamma_{p,l,p',l'}:=~\gamma_S-\gamma_{p',l'}=~2\gamma_{p,l}-~\gamma_{p',l'}$, where $\gamma_S$ is the Gouy phase factor of the crystal polarisation, $\gamma_{p',l'}$ is the Gouy phase factor of the corresponding harmonic mode and $\Delta\gamma_{p,l,p',l'}>0$, always. This decay can be slowed down (or amplified) by a wavevector mismatch which is, however, also the same for each term in the crystal polarisation. It can thus only be properly used for one of the harmonic modes. Hence, good phase matching is either achieved for each harmonic mode when approaching the limit of plane waves because the Gouy phases are then negligible. Or it can only be achieved for one of the harmonic modes, at most.

The same applies to the HG basis with $\Delta\gamma_{m,n,m',n'}=~2\gamma_{m,n}-~\!\gamma_{m',n'}$. In the LG basis, however, the selection of excited harmonic modes leads to $\Delta\gamma_{p,l,p',l'}$ values which decrease in steps of 4 down to the minimum of 1 for $p'=0,2,\dots,2p$. In contrast, the HG basis also allows for other values inbetween.

\section{Change in power and phase difference}\label{sec:changeInPowerAndPhase}
Comparing Eqs.\ \ref{DEQ:discreteHelmholtzWithModes} and \ref{DEQ:sourceTermComplete} gives
\begin{equation}
c_{2m,l'}(z)=c_{p',l'}(z)=i\mathcal{C}_{p,l}(z)c'_{p',l'}e^{i\Delta\gamma_{p,l,p',l'}\Psi(z)}e^{i\Delta kz}
\end{equation}
such that the change in power for each mode in the harmonic field at $z$ is given by
\begin{align}
\begin{split}
\Delta P^h_{p',l'}(z)&\propto\left|a_{p',l'}(z)\right|^2-\left|a_{p',l'}(z-\Delta z)\right|^2\\
&\propto\left|\mathcal{C}_{p,l}(z)c'_{p',l'}\right|^2+2\left|a_{p',l'}(z-\Delta z)\mathcal{C}_{p,l}(z)c'_{p',l'}\right|\times\cos\big(\Delta\alpha_{p',l'}(z)\big)
\end{split} 
\label{eqn:changeInPower}
\end{align}
with the phase difference
\begin{align}
\Delta\alpha_{p',l'}(z)&=\alpha_{a,p',l'}(z-\Delta z)-\frac{\pi}{2}-\Delta\gamma_{p,l,p',l'}\Psi(z)-\Delta kz
\label{eqn:phaseDifference}
\end{align}
where $\alpha_{a,p',l'}$ is the phase of $a_{p',l'}$ and $\pi/2$ corresponds to the factor $i$. $\Delta\alpha_{p',l'}(z)$ will be called Gouy phase difference for $\Delta k=0$.  

Let us consider the second expression for $\Delta P^h_{p',l'}(z)$. The first term is only important at $z_0$ when there is no harmonic field yet. It quickly becomes negligible compared to the second term as soon as the conversion begins, especially for a higher $z$-resolution (smaller $\Delta z$) because it is quadratic in $\Delta z$ while the second term is linear in $\Delta z$. The second term shows that the change in the power of each mode in the harmonic field depends on four parameters:
\begin{enumerate}
	\item The contribution of the respective mode to the crystal polarisation $c'_{p',l'}$ which is constant throughout the crystal. 
	\item The amplitude of the crystal polarisation $\mathcal{C}_{p,l}(z)$. This parameter includes the influence of the pump intensity. For a set pump power, $\mathcal{C}_{p,l}(z)$ is larger for more intense pump modes (the latter generally implies a smaller mode order).  $\mathcal{C}_{p,l}(z)$, additionally, decreases with increasing pump beam radius. Both effects are shown in Fig.\ \ref{fig:crystalPolarisationComparison} for the later simulated pump modes $\text{LG}_{0,0}$ and $\text{LG}_{3,3}$ depending on the focusing parameter $\xi=~\!L/(2z_R)$, where $L$ is the geometrical length of the crystal. Larger focusing parameters imply a smaller waist size and, thus, a higher maximum intensity but also a larger beam divergence. Hence, $\mathcal{C}_{p,l}(z)$ or $|\mathcal{C}_{p,l}(z)|^2$ changes more extremely for larger focusing parameters. 
	
	Comparing Eqs.\ \ref{DEQ:initialForm} and \ref{DEQ:sourceTermComplete} shows that the effective nonlinearity can be transformed into a pump mode-dependent effective nonlinearity 
	\begin{equation}
	d_{p,l}(z):=d_\text{eff}\mathcal{C}_{p,l}(z) \text{.}
	\label{eqn:modeDependentNonlinearity}
	\end{equation}
	The ratio $d_{p1,l1}/d_{p2,l2}$ for two different pump modes, e.g.\ $d_{0,0}/d_{3,3}=2.64$, is then $z$-, $\xi$- and medium-independent ($d_\text{eff}$ cancels out) and directly quantifies the difference of the two pump modes in their intensity distributions as well as in their interaction with the nonlinear medium. 
	\begin{figure}[htbp]
		\centering
		\hspace*{-0.5cm}\includegraphics[trim=2.8cm 10.5cm 4.4cm 10.7cm,clip,width=8cm]{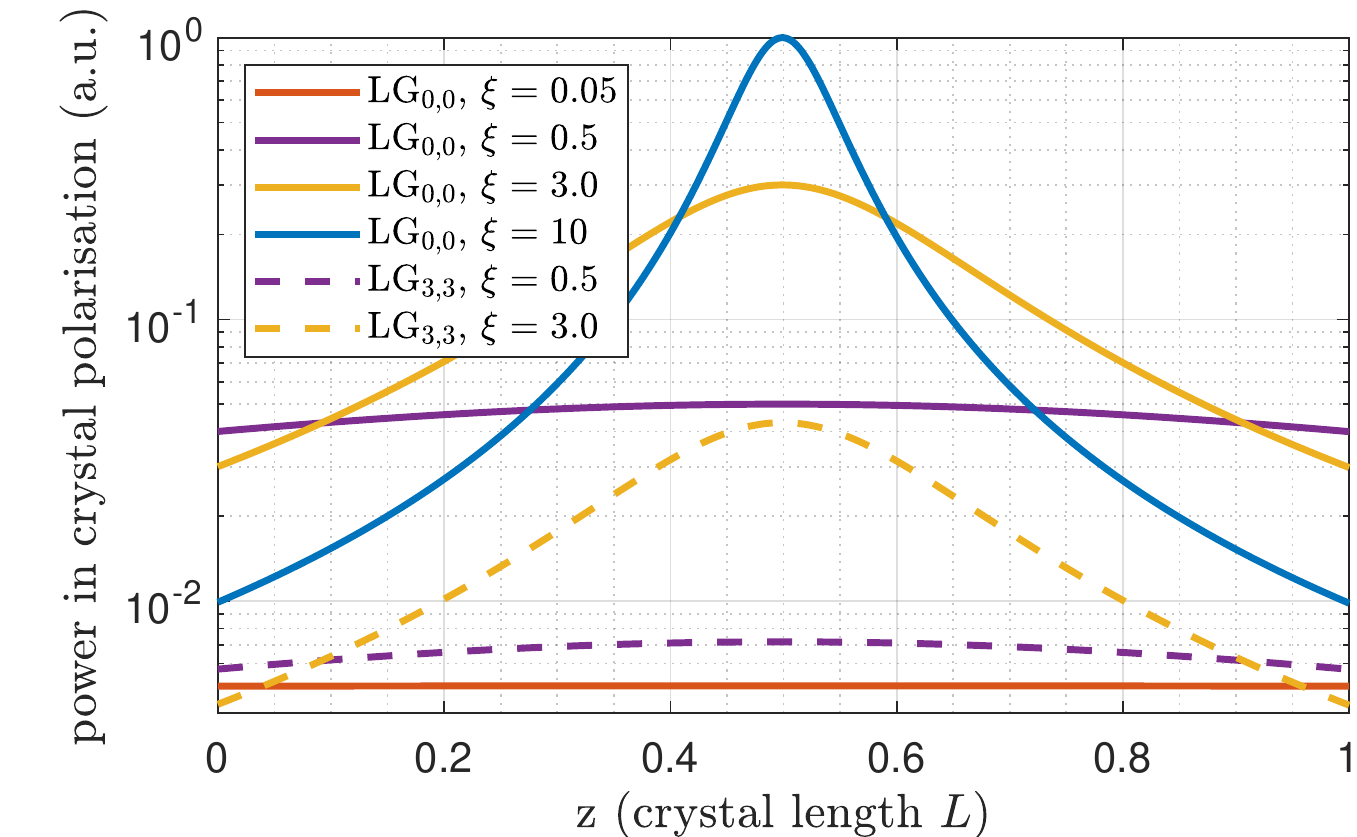}
		\caption{Normalised evolution of the power contained in the crystal polarisation $|\mathcal{C}_{p,l}(z)|^2$ for different focusing parameters $\xi$ along the crystal on a logarithmic scale. The waist of the pump field is located at the centre of the crystal. The curves evolve in accordance with the pump intensities.}
		\label{fig:crystalPolarisationComparison}
	\end{figure}
	\item The already generated power in the respective mode itself in terms of $|a_{p',l'}(z-\Delta z)|$ such that the power will change more if more power has already been generated. 
	\item The phase difference $\Delta\alpha_{p',l'}(z)$ between $a_{p',l'}(z-~\!\Delta z)$ and $c_{p',l'}(z)$. According to the cosine in Eq.\ \ref{eqn:changeInPower}, the phase difference will cause the power to decrease whenever $|\Delta\alpha_{p',l'}(z)|>\pi/2$ (and if the second term dominates the first one at this point). As expected, the evolution of the phase difference depends on the difference in the Gouy phase factors $\Delta\gamma_{p,l,p',l'}$ and on the wavevector mismatch $\Delta k$. The phase term $\pi/2$ is imprinted on the phase of $a_{p',l'}$ at $z_0$ and irrelevant for the phase matching. The same would hold for $\beta\neq 0$ (see Eq.\ \ref{eqn:LGmode}). Furthermore, the phase of $a_{p',l'}$ adjusts to the phase of $c_{p',l'}$ via the interference in each iteration step and the evolution of the phase difference highly depends on how efficient this adjustment works. This is mainly determined by the slope of the phase of $c_{p',l'}(z)$. If this phase changes rapidly, the phase of $a_{p',l'}$ can not adjust quickly enough. Another influence is the ratio $|c_{p',l'}(z)|/|a_{p',l'}(z-\Delta z)|$; the larger this ratio is, the more does $c_{p',l'}(z)$ affect the phase of $a_{p',l'}(z-\Delta z)$.
\end{enumerate}

How the SHG depends on these four parameters will be illustrated in the following via numerical simulations using Eq.\ \ref{DEQ:discreteHelmholtzWithModes}.

\section{$\text{LG}_{0,0}$ in single-pass SHG}\label{sec:LG00}
\begin{figure}[htbp]
	\centering
	\begin{tabular}{cc}
		\captionsetup[subfigure]{oneside,margin={0cm,0cm}}
		\subfloat{\hspace*{-0.4cm}\includegraphics[trim=3cm 9.35cm 3cm 9.45cm,clip,width=0.5\linewidth]{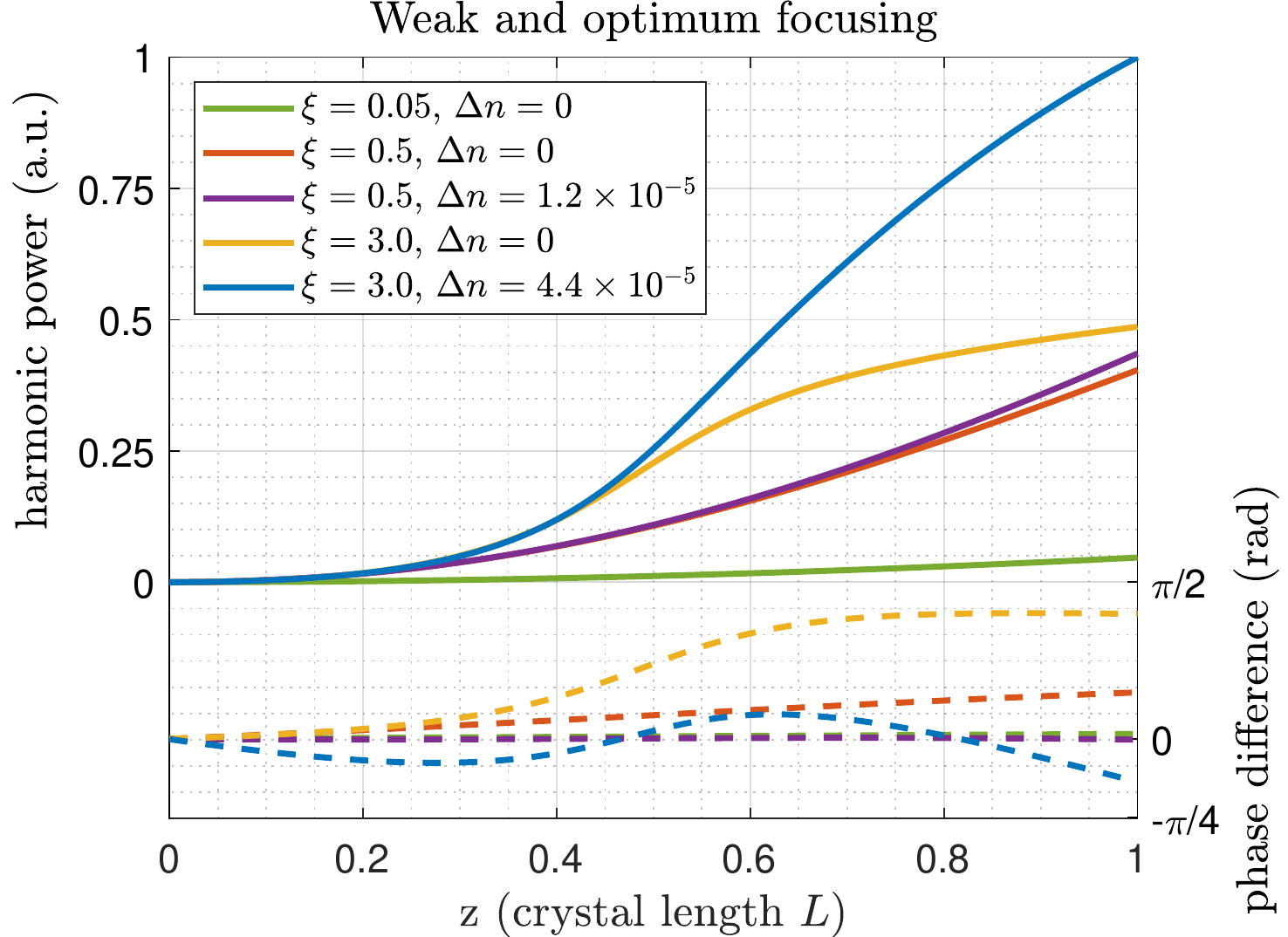}}
		\captionsetup[subfigure]{oneside,margin={0cm,0cm}}
		\subfloat{\hspace*{0cm}\includegraphics[trim=3cm 9.35cm 3cm 9.45cm,clip,width=0.5\linewidth]{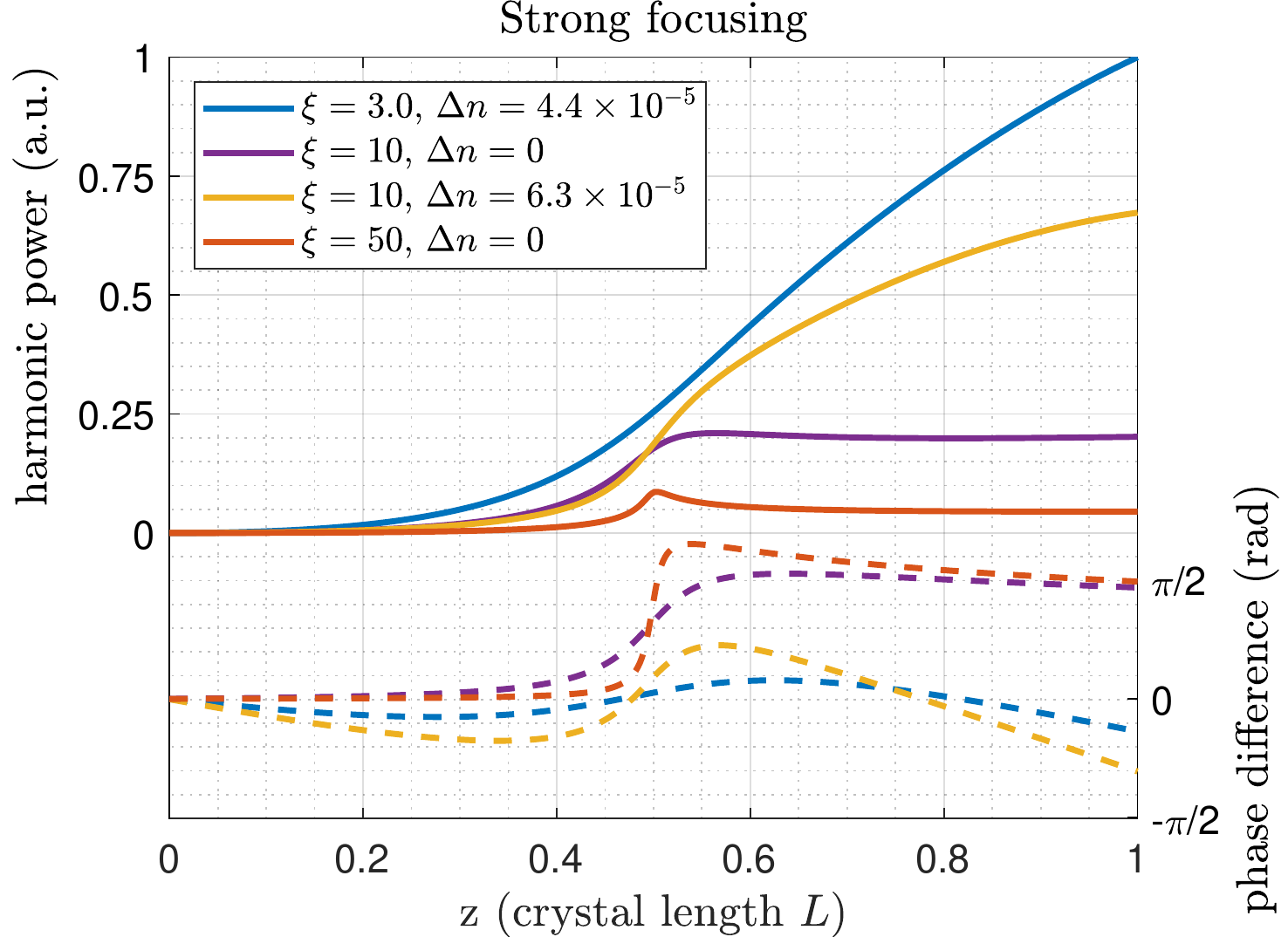}}
	\end{tabular}
	\caption{Numerical simulation of the SHG pumped by the $\text{LG}_{0,0}$ mode for different focusing parameters $\xi$ and wavevector mismatches $\Delta n$. The waist is located at the centre of the crystal.  The solid lines refer to the left axis and show the accumulated harmonic power. The dashed lines refer to the right axis and illustrate the phase difference between the harmonic field and the crystal polarisation for the harmonic $\text{LG}_{0,0}$ mode.}
	\label{fig:LG00simulation}
\end{figure}

The $\text{LG}_{0,0}$ mode is the lowest order ($g_{0,0}=0$) LG mode, shows the highest intensity for a set optical power and should therefore generally achieve the highest conversion efficiency. If pumped by this $p=0$ mode, the crystal polarisation only excites the harmonic $\text{LG}_{0,0}$ mode with reduced waist size and $\Delta\gamma_{p,l,p',l'}=~\Delta\gamma_{0,0,0,0}=~1$. Thus, the phase difference evolves with $\Delta\gamma_{p,l,p',l'}\Psi(z)=~\!\text{atan}(z/z_R)$ (see Eq.\ \ref{eqn:phaseDifference}). Fig.\ \ref{fig:LG00simulation} shows how the conversion evolves along the crystal dependend on the focusing parameter $\xi$ and regarding the accumulated harmonic power and phase difference. It is split up into the regime of weak to optimum focusing (optimum focusing being defined as $\xi=2.84$ according to \cite{BoydKleinmann} for highest conversion efficiency) and the regime of strong focusing. In each case, the waist is located at the centre of the crystal and the phase difference starts at zero because the harmonic field is equal to the crystal polarisation after the first iteration step in Eq.\ \ref{DEQ:firstStep}. 

Please note that the simulation produces the same relative results for any values $\lambda_1$, $n$ and $d_\text{eff}$ and any pump power if the pump field is treated as undepleted. Furthermore, the ratios of the final harmonic power in the different cases are equal to the ratios of the corresponding conversion efficiencies because the same pump power is assumed for all simulated cases.  

For $\xi=0.05$, the focusing is weak enough to be considered close to the limit of plane waves in the sense that the Gouy phase difference remains negligibly small. Hence, the increase in harmonic power is not affected by the Gouy phase difference. Still, the conversion efficiency is only at \SI{5}{\percent} of the most efficient simulated case ($\xi=3$) because the pump intensity, or $\mathcal{C}_{0,0}(z)$, is comparably small. Please note further that the steady increase in the harmonic power dominates the changes in $\mathcal{C}_{0,0}(z)$ and results in a steady increase in the slope of the harmonic power according to Eq.\ \ref{eqn:changeInPower}.

For $\xi=0.5$, the focusing is still relatively weak. In the range $-L/2\leq z\leq L/2$ (the $z$-range used in the simulation in contrast to the $x$-axis in Fig.\ \ref{fig:LG00simulation}), $\text{atan}(z/z_R)$ covers a span of almost $\pi/3$. However, due to the adjustment of the phase of the harmonic field to the phase of the crystal polarisation, the Gouy phase difference remains below $\pi/4$. It does not significantly affect the evolution of the harmonic power. This becomes apparent when the optimum wavevector mismatch is included: $\text{atan}(z/z_R)$ evolves approximately linearly within the crystal such that the wavevector mismatch can properly compensate for it. But the improvement on the harmonic power is small compared to $\xi=3\ \text{or}\ 10$. The higher pump intensity results in a sevenfold increase of the conversion efficiency compared to $\xi=0.05$.

For $\xi=3$, the harmonic power builds up with a larger slope along roughly two third of the crystal length compared to $\xi=0.5$ due to the higher pump intensity. At the same time, the Gouy phase difference can not be neglected anymore and continuously reduces the slope of the harmonic power. Here, $\text{atan}(z/z_R)$ covers a span of about $4\pi/5$ in the range $-L/2\leq z\leq L/2$ and the Gouy phase difference remains below $\pi/2$ due to the phase adjustment of the harmonic field. With this Gouy phase difference, the final harmonic power is only slightly higher than for $\xi=0.5$. In this focusing regime, the optimum wavevector mismatch can not keep the phase difference close to zero throughout the crystal because the Gouy phase difference does not evolve linearly anymore. Nevertheless, the wavevector mismatch can improve the conversion efficiency by a factor of about 2.

For $\xi=10$, the focusing is strong enough to cause a significant pump beam divergence. This becomes apparent in the first crystal half where the low pump intensity does not allow the harmonic power to reach higher values than for $\xi=3$ despite the small Gouy phase difference. Approaching the waist, the pump intensity reaches higher values than for $\xi=3$ and the harmonic power increases with larger slope. At the same time, the Gouy phase difference starts to significantly reduce the slope of the harmonic power around the waist. While $\text{atan}(z/z_R)$ almost covers its complete span of $\pi$ in the range $-L/2\leq z\leq L/2$, the Gouy phase difference reaches and remains close to $\pi/2$. This causes the harmonic power to remain roughly constant in the second half of the crystal on a level which is even lower than for $\xi=0.5$. The optimum wavevector mismatch can triple the final harmonic power; but since the evolution of the Gouy phase difference deviates even more from a linear increase than for $\xi=3$, this compensation is clearly less efficient and the conversion efficiency is smaller than for $\xi=3$ including the respective optimum wavevector mismatch. Evidently, there is an optimum focusing parameter regarding the conversion efficiency with the best compromise between pump intensity and achievable phase matching. $\xi_\text{opt}=2.84$ was derived in \cite{BoydKleinmann} including a $\xi$-dependent optimum wavevector mismatch.

Please note that the nTADs of the harmonic field and crystal polarisation differ by not even \SI{1}{ppm} within the crystal for $\Delta n=~6.3\times 10^{-5}$ and can therefore be treated as identical (concerning the remark in brackets before Eq.\ \ref{DEQ:discreteHelmholtzWithModes}).

For $\xi=50$, the effect of the pump beam divergence is clearly larger than for $\xi=10$. Around the waist, where the pump intensity reaches maximum values, the harmonic power briefly increases with the maximum slope of the simulated cases until the Gouy phase difference approaches $\pi/2$. In this case, the Gouy phase difference even exceeds $\pi/2$ due to the rapid evolution of $\text{atan}(z/z_R)$ such that the harmonic power decreases. For $z>0.05L$ (referring to the $x$-axis), $\text{atan}(z/z_R)$ is already approximately constant at $\pi/2$. In this $z$-range, only the phase of the harmonic field changes due to the phase adjustment; hence, the Gouy phase difference decreases. The conversion efficiency is at the same level as for $\xi=0.05$.

\begin{figure}[htbp]
	\centering
	\hspace*{-0.2cm}\includegraphics[trim=3cm 11.05cm 3.3cm 11.4cm,clip,width=8cm]{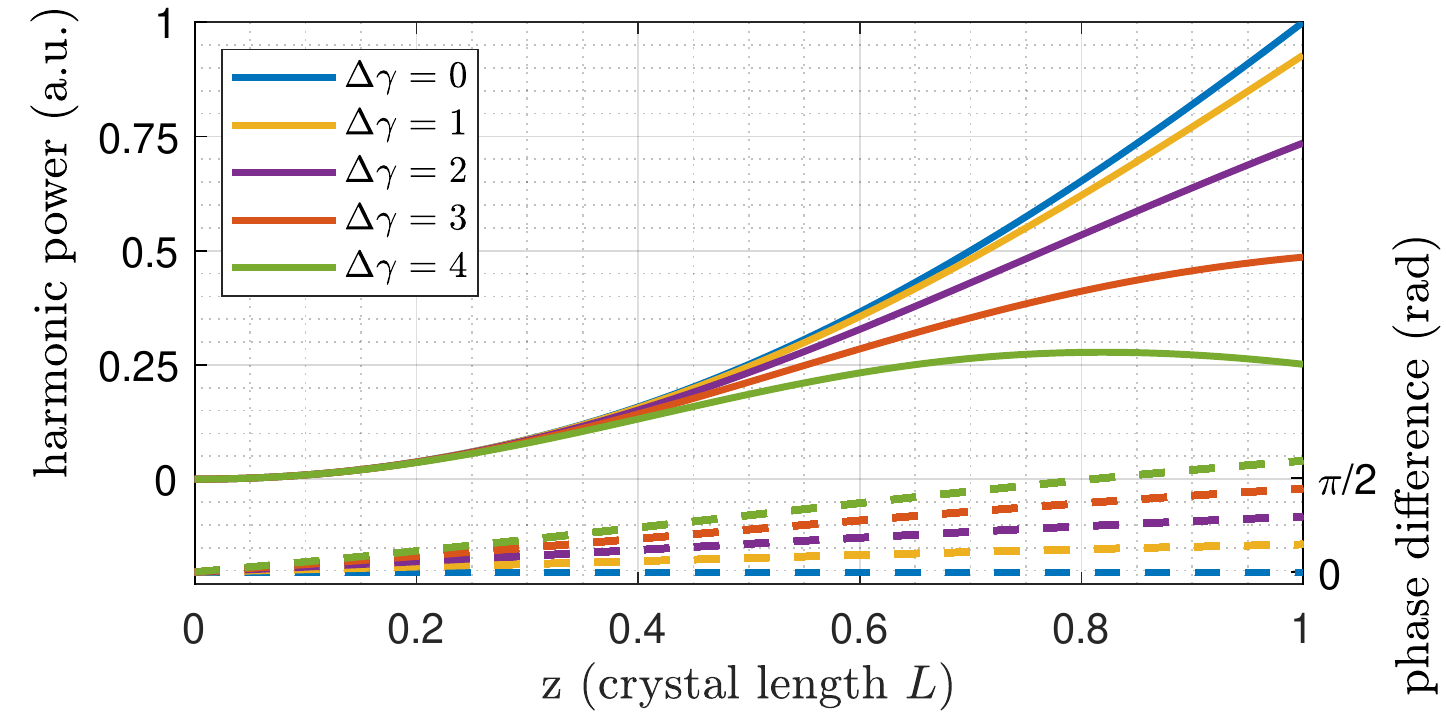}
	\caption{Demonstration of the influence of $\Delta\gamma_{p,l,p',l'}$, here artificially varied for $\xi=0.5$ and $\Delta n=0$. The solid lines refer to the left axis and show the accumulated harmonic power. The dashed lines refer to the right axis and illustrate the Gouy phase difference.}
	\label{fig:LG00artificially}
\end{figure}

Finally, Fig.\ \ref{fig:LG00artificially} demonstrates how a step-wise increase in $\Delta\gamma_{p,l,p',l'}$ deteriorates the power build-up in the harmonic field. For this demonstration, $\Delta\gamma_{0,0,0,0}$ is varied artificially for $\xi=0.5$, even though $\Delta\gamma_{0,0,0,0}\neq 1$ is not physical. As expected, a higher difference in the Gouy phase factors increases the slope of the Gouy phase difference, whereas the unphysical case of $\Delta\gamma_{p,l,p',l'}=0$ behaves like the limit of plane waves with no Gouy phase difference at all.

\section{$\text{LG}_{3,3}$ in single-pass SHG}\label{sec:LG33}
\begin{figure*}[!]
	\centering
	\begin{tabular}{cc}
		\captionsetup[subfigure]{oneside,margin={0cm,0cm}}
		\subfloat{\hspace*{-0.4cm}\includegraphics[trim=3cm 9.35cm 3cm 9.45cm,clip,width=0.5\linewidth]{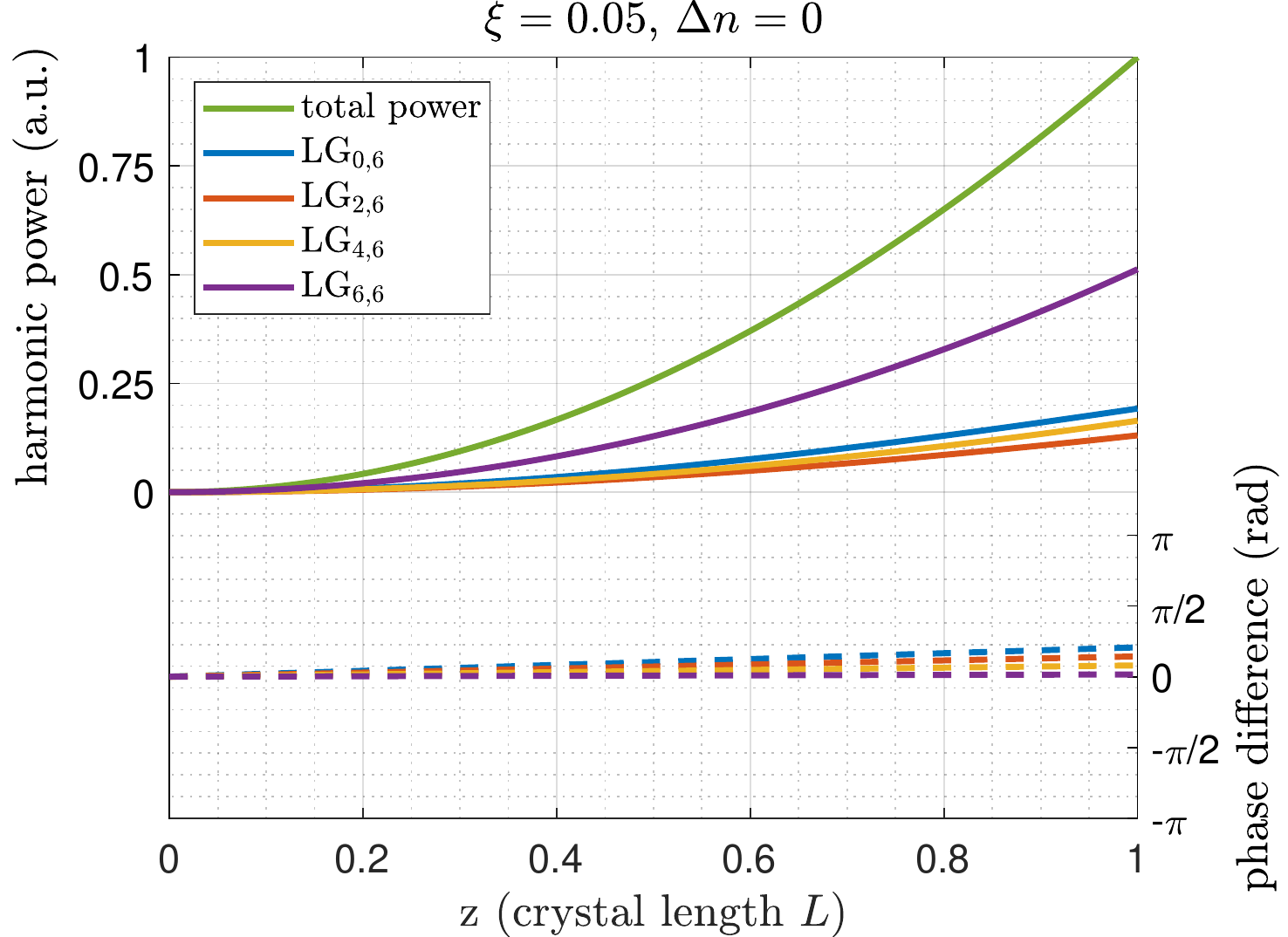}}
		\captionsetup[subfigure]{oneside,margin={0cm,0cm}}
		\subfloat{\hspace*{0cm}\includegraphics[trim=3cm 9.35cm 3cm 9.45cm,clip,width=0.5\linewidth]{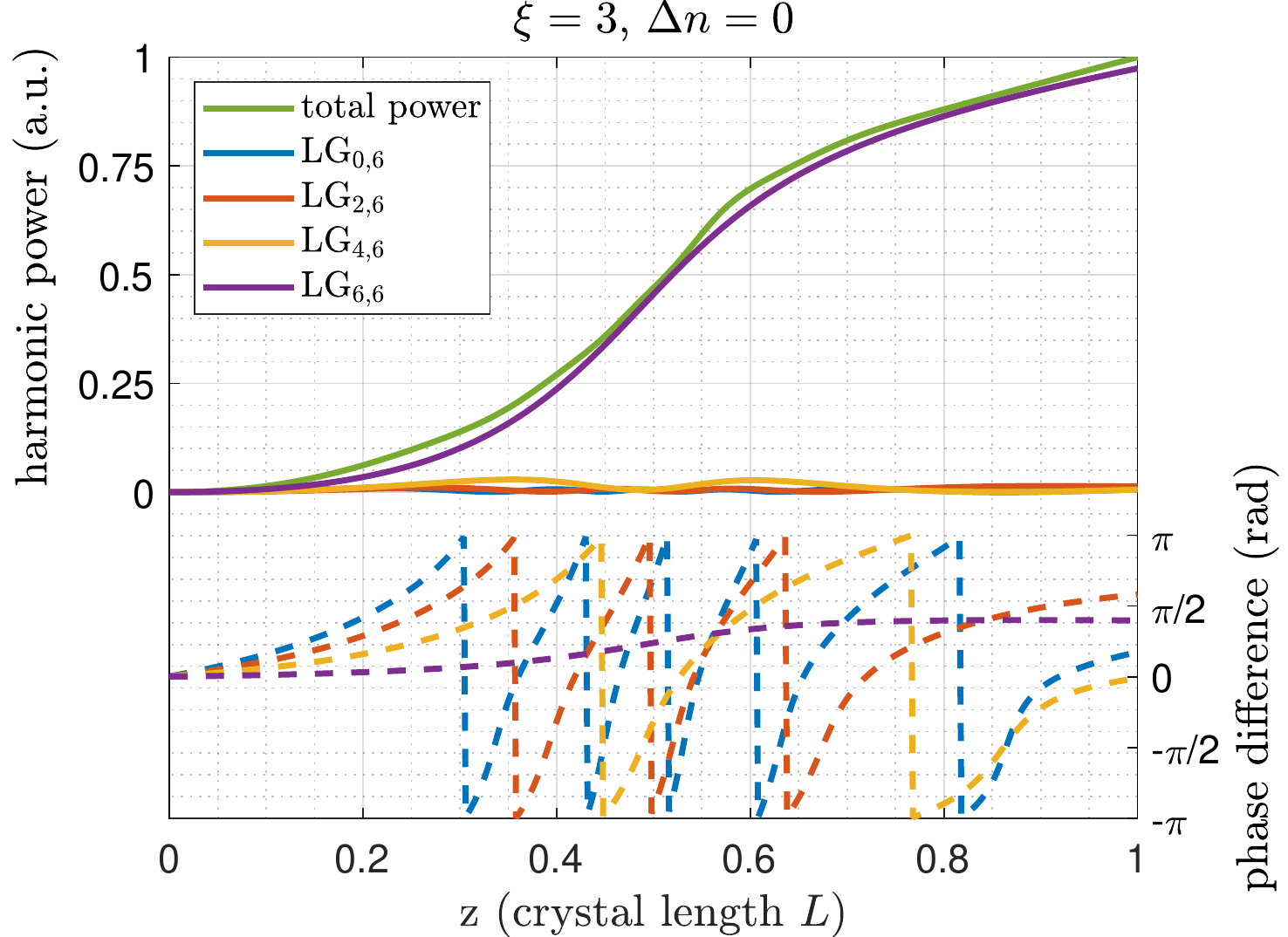}}\\ & \\ 
		\captionsetup[subfigure]{oneside,margin={0cm,0cm}}
		\subfloat{\hspace*{-0.4cm}\includegraphics[trim=3cm 9.35cm 3cm 9.45cm,clip,width=0.5\linewidth]{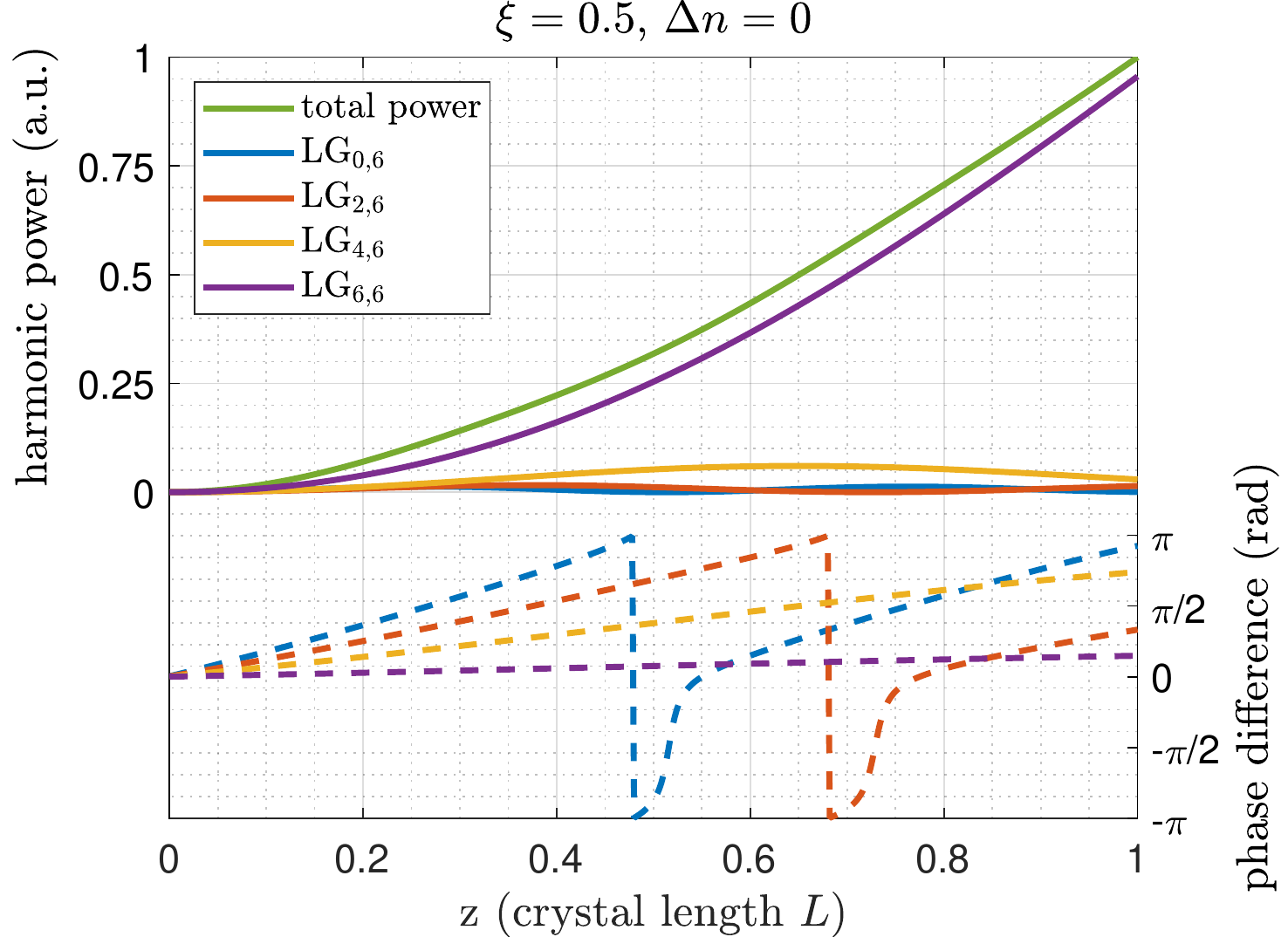}}
		\captionsetup[subfigure]{oneside,margin={0cm,0cm}}
		\subfloat{\hspace*{0cm}\includegraphics[trim=3cm 9.35cm 3cm 9.45cm,clip,width=0.5\linewidth]{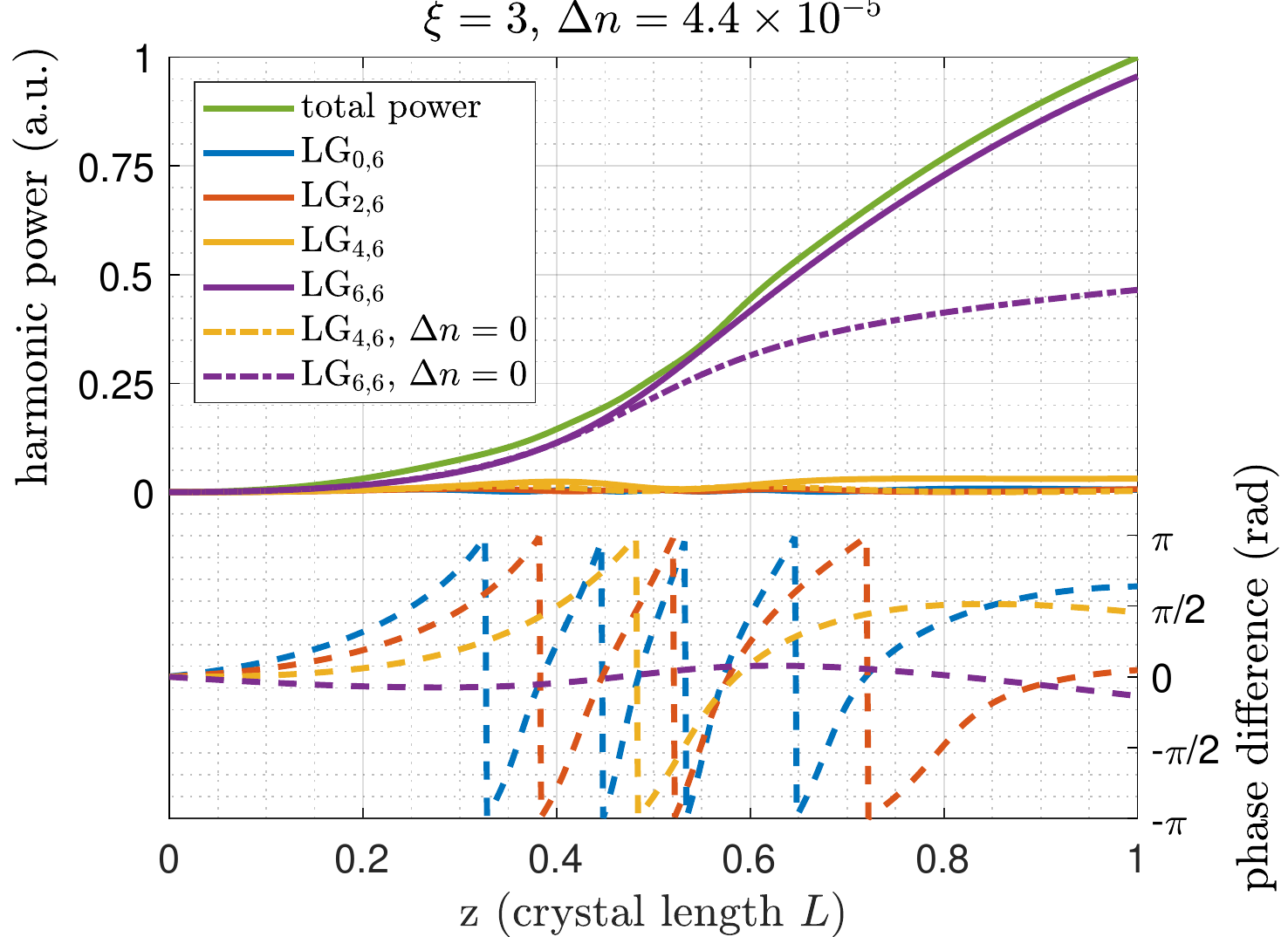}}\\ & \\ 
		\captionsetup[subfigure]{oneside,margin={0cm,0cm}}
		\subfloat{\hspace*{-0.4cm}\includegraphics[trim=3cm 9.35cm 3cm 9.45cm,clip,width=0.5\linewidth]{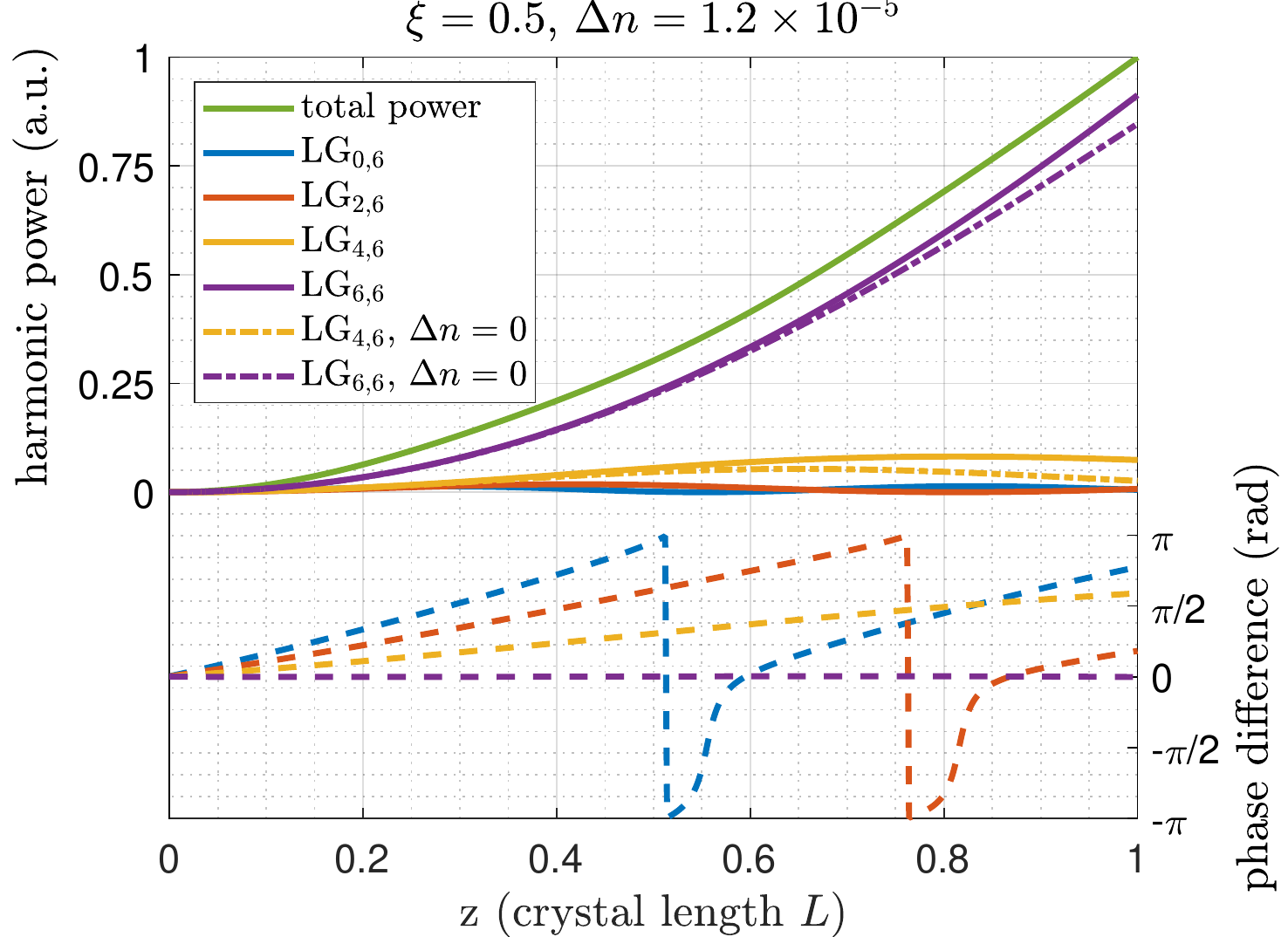}}
		\captionsetup[subfigure]{oneside,margin={0cm,0cm}}
		\subfloat{\hspace*{0cm}\includegraphics[trim=3cm 9.35cm 3cm 9.45cm,clip,width=0.5\linewidth]{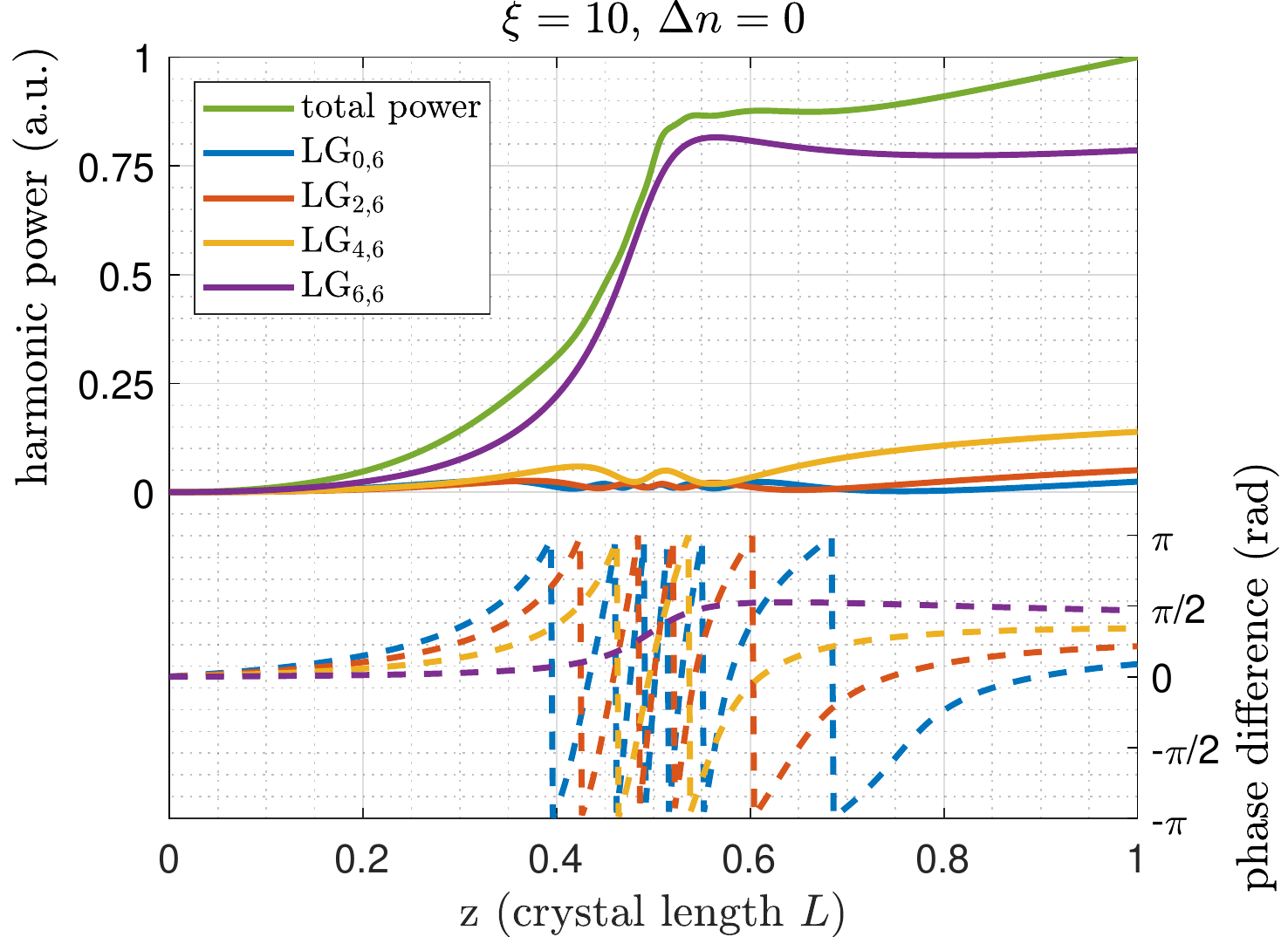}}\\
	\end{tabular}
	\caption{Numerical simulation of SHG pumped by the $\text{LG}_{3,3}$ mode for different focusing parameters $\xi$ and wavevector mismatches $\Delta n$. The waist is located at the centre of the crystal. The solid lines refer to the left axis and show the accumulated harmonic power. The dashed lines refer to the right axis and illustrate the phase difference between the respective mode in the harmonic field and the crystal polarisation.}
	\label{fig:LG33simulation_weak}
\end{figure*}

The $\text{LG}_{3,3}$ mode is a higher-order pump mode ($g_{3,3}=~\! 9$) with $p>0$ and should fundamentally show a much lower conversion efficiency than the $\text{LG}_{0,0}$ mode due to the more uniform intensity distribution. The latter implies $\mathcal{C}_{0,0}(z)>\mathcal{C}_{3,3}(z)$ for equal focusing parameters. When pumped by the $\text{LG}_{3,3}$ mode, the crystal polarisation locally excites a superposition of harmonic modes with reduced waist size according to Tab.\ \ref{tab:sourceTermLG33}. 
\begin{table}[h!]
	\centering
	\caption{\bf Composition of crystal polarisation excited by the $\text{LG}_{3,3}$ mode}
	\begin{tabular}{ccc}
		\hline\vspace*{-0.6cm}&&\\\vspace*{0.1cm}
		$p',l'$ &\qquad $|c'_{p',l'}|^2$ &\quad $\Delta\gamma_{3,3,p',l'}$ \\
		\hline\vspace*{-0.6cm} &&\\
		0,6 &\qquad 0.2129 &\quad 13 \\
		2,6 &\qquad 0.1342 &\quad 9 \\
		4,6 &\qquad 0.1610 &\quad 5 \\
		6,6 &\qquad 0.4919 &\quad 1 \\
		\hline
	\end{tabular}
	\label{tab:sourceTermLG33}
\end{table}
Fig.\ \ref{fig:LG33simulation_weak} shows how the power of these modes in the harmonic field, the total harmonic power as well as their respective phase differences evolve along the crystal for the same focusing parameters as in the case of the $\text{LG}_{0,0}$ mode (except for $\xi=50$). For $\xi=0.5\ \text{and}\ 3$, they also illustrate the influence of the same wavevector mismatches for comparison. 

Please note that $\Delta\gamma_{0,0,0,0}=\Delta\gamma_{3,3,6,6}$. Hence, the Gouy phase difference of the harmonic $\text{LG}_{6,6}$ mode, when pumped by the $\text{LG}_{3,3}$ mode, behaves identically to the case of the harmonic $\text{LG}_{0,0}$ mode, when pumped by the $\text{LG}_{0,0}$ mode, for equal focusing parameters. The power in these two cases only evolves qualitatively identically due to the different pump intensities. This correlation holds for any two pairs of pump and harmonic mode with equal $\Delta\gamma_{p,l,p',l'}$ value.
 
For $\xi=0.05$, the increase in harmonic power is hardly affected for any of the harmonic modes, even though the influence of the different $\Delta\gamma_{3,3,p',l'}$ values can be seen in the different slopes of the Gouy phase differences. Hence, their final contributions to the total harmonic power are roughly equal to the respective $|c'_{p',l'}|^2$ values such that the harmonic field remains in roughly the same mode composition as the crystal polarisation.

For $\xi=0.5$, the conversion into the harmonic $\text{LG}_{6,6}$ mode is still hardly affected by the Gouy phase difference. However, the stepwise increase of 4 in the values $\Delta\gamma_{3,3,6,6}$ to $\Delta\gamma_{3,3,0,6}$ causes the Gouy phase difference of each of the other harmonic modes to (repeatedly) reach values where destructive interference dominates ($|\Delta\alpha_{p',l'}|>\pi/2$). Hence, the conversion efficiencies of the harmonic modes $\text{LG}_{0,6}$, $\text{LG}_{2,6}$ and $\text{LG}_{4,6}$ are jointly almost negligible compared to the $\text{LG}_{6,6}$ mode such that the latter finally dominates the harmonic field with a contribution of about \SI{95}{\percent}. 

Please note that the Gouy phase difference changes more rapidly after reaching $-\pi$ because the power in the respective mode is close to zero at this point. Hence, the influence of the crystal polarisation on the phase of this mode in the harmonic field is larger. 

In general, a wavevector mismatch reduces the slope of the phase difference for each harmonic mode and, again, cases with equal $\Delta\gamma_{p,l,p',l'}$ values benefit (or suffer) in the same manner. For $\xi=0.5$, the improvement in the conversion efficiency of the harmonic $\text{LG}_{6,6}$ mode via the same wavevector mismatch as before in the $\text{LG}_{0,0}$ case is small because the Gouy phase difference is not significant. As the other extreme, the slopes of the Gouy phase differences of the $\text{LG}_{0,6}$ and $\text{LG}_{2,6}$ mode are not sufficiently affected by the wavevector mismatch to significantly increase their conversion efficiency. On the other hand, the $\text{LG}_{4,6}$ mode benefits the most from the wavevector mismatch (in relative terms) because its Gouy phase difference evolves up to an average region around $\pi/2$. Hence, its contribution to the final harmonic field increases by about \SI{5}{\percent} such that the contribution of the $\text{LG}_{6,6}$ mode decreases to about \SI{90}{\percent}. Interestingly, the optimum wavevector mismatch now depends on the goal of the SHG. If a high total harmonic power is desired, the optimum wavevector mismatch will be larger than simulated here to still increase the benefit for the $\text{LG}_{4,6}$ mode while not significantly deteriorating the power increase of the $\text{LG}_{6,6}$ mode. If a high purity in terms of $\text{LG}_{6,6}$ is desired, no wavevector mismatch or even a mismatch with opposite sign will be ideal to achieve a higher suppression of the other modes.  

For $\xi=3$, the $\text{LG}_{3,3}$ mode achieves the highest of the simulated total conversion efficiencies (corresponding to the final total harmonic power). Without a wavevector mismatch, the contribution of the harmonic $\text{LG}_{6,6}$ mode to the final harmonic field is above \SI{95}{\percent} because the suppression of the other harmonic modes in relation to the $\text{LG}_{6,6}$ mode is stronger than for $\xi=0.5$. In this focusing regime, the $\text{LG}_{4,6}$ mode again benefits more from the simulated wavevector mismatch than the $\text{LG}_{6,6}$ mode (in relative terms) such that the latter's contribution to the final harmonic field is reduced to \SI{95}{\percent}.

For $\xi=10$, the performance of the $\text{LG}_{0,6}$, $\text{LG}_{2,6}$ and $\text{LG}_{4,6}$ mode does not significantly change compared to the other simulated focusing cases apart from different patterns of power modulation. Due to the same reason as for the $\text{LG}_{0,0}$ mode for this focusing regime, the generated power in the $\text{LG}_{6,6}$ mode is, however, clearly reduced compared to $\xi=\ $\numlist{0.5;3}. Hence, its contribution to the final harmonic field is only at about \SI{80}{\percent}.

\section{Comparison of $\text{LG}_{0,0}$ and $\text{LG}_{3,3}$ in single-pass SHG}
\begin{figure}[htbp]
	\centering
	\hspace*{-0.4cm}\includegraphics[trim=3.2cm 10.2cm 4.2cm 10.5cm,clip,width=8cm]{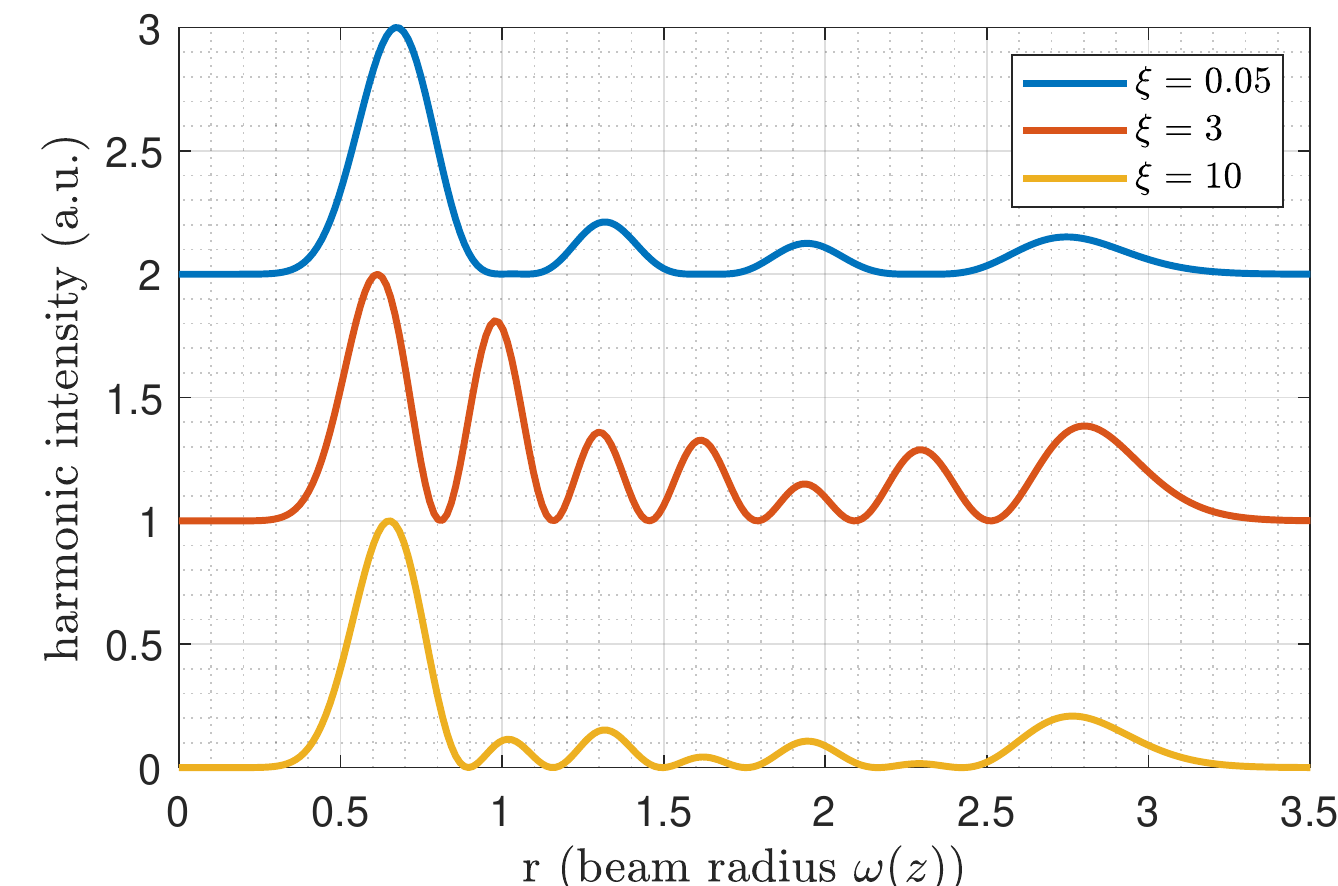}
	\caption{Normalised final harmonic intensity distribution in the far-field ($z=100z_R$) for different focusing parameters $\xi$ when pumped by the $\text{LG}_{3,3}$ mode.}
	\label{fig:finalHarmonicIntensity}
\end{figure}

First, we discuss the harmonic output field and its dependency on the focusing. When the SHG is pumped by the $\text{LG}_{0,0}$ mode, only the harmonic $\text{LG}_{0,0}$ mode is excited such that the harmonic output field does not depend on the focusing in terms of its mode composition (in contrast to its power). This is different when the SHG is pumped by the $\text{LG}_{3,3}$ mode. Here, four harmonic modes are excited and each of them shows an individual focusing-dependent conversion efficiency. Hence, the contributions of the four harmonic modes to the final harmonic field change dependent on the focusing parameter such that the final harmonic mode composition and intensity distribution change as well. This is shown in Fig.\ \ref{fig:finalHarmonicIntensity} for the far-field. As expected, the harmonic intensity distribution for $\xi=0.05$ is equal to the squared intensity distribution of the $\text{LG}_{3,3}$ mode. For $\xi=3$, it is dominated by the $\text{LG}_{6,6}$ mode and for $\xi=10$, the peak structure of the $\xi=0.05$ case is dominating because the conversion into the $\text{LG}_{6,6}$ mode is less efficient again.

\begin{figure}[htbp]
	\centering
	\hspace*{-0.4cm}\includegraphics[trim=3.2cm 11.2cm 4.2cm 11.4cm,clip,width=8cm]{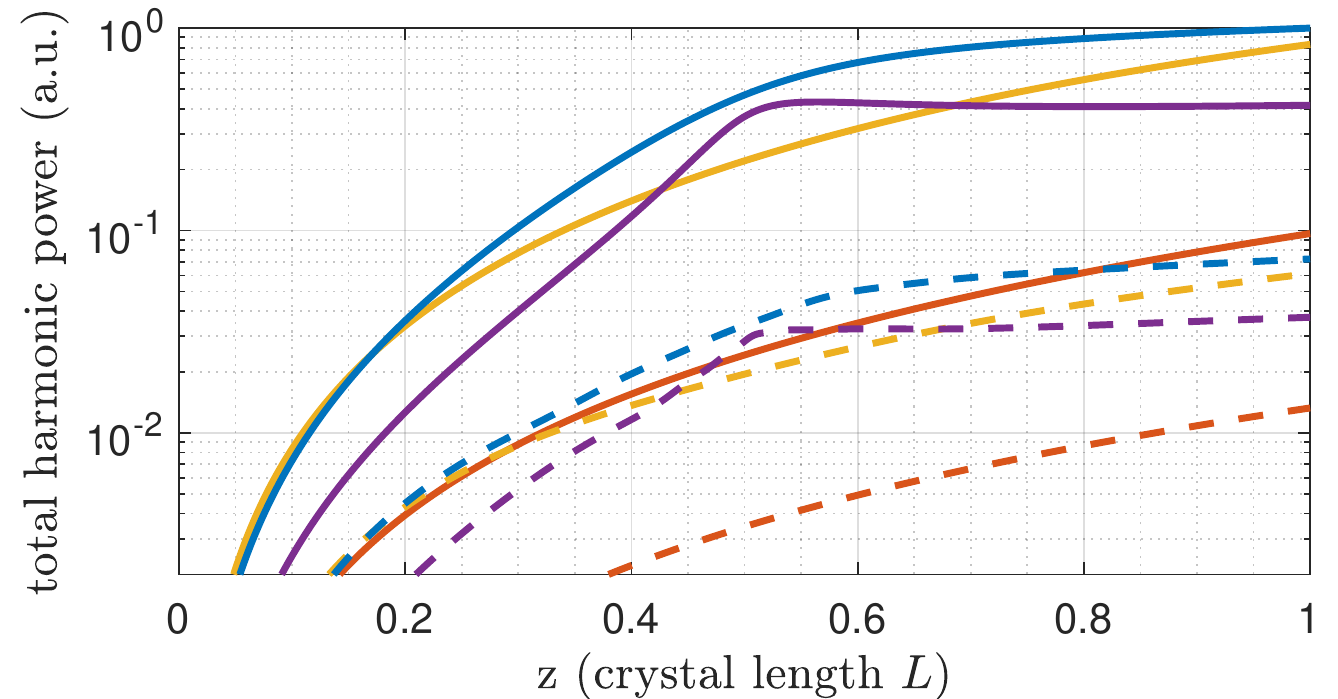}
	\caption{Comparison of the accumulated total harmonic power for different focusing parameters $\xi$ and no wavevector mismatch ($\Delta n=0$). The solid and dashed lines refer to the $\text{LG}_{0,0}$ and $\text{LG}_{3,3}$ case, respectively. Colour code: orange $\leftrightarrow$ $\xi=0.05$, yellow $\leftrightarrow$ $\xi=0.5$, blue $\leftrightarrow$ $\xi=3$ and purple $\leftrightarrow$ $\xi=10$.}
	\label{fig:powerComparison}
\end{figure}

Secondly, we compare the total conversion efficiencies (in the case of $\text{LG}_{3,3}$ referring to the total harmonic power). Fig.\ \ref{fig:powerComparison} shows the evolution of the accumulated total harmonic power when pumped by the fundamental $\text{LG}_{0,0}$ and $\text{LG}_{3,3}$ mode for different focusing parameters and without a wavevector mismatch. At the beginning of the crystal, the ratio between curves of equal focusing parameter is equal to $(d_{0,0}/d_{3,3})^2=6.97$ (see Eq.\ \ref{eqn:modeDependentNonlinearity}). However, since the power in the harmonic modes $\text{LG}_{0,6}$, $\text{LG}_{2,6}$ and $\text{LG}_{4,6}$ becomes ever more negligible along the crystal for the majority of the simulated focusing parameters, this ratio changes. If the harmonic $\text{LG}_{6,6}$ mode is finally dominating, it approaches $(d_{0,0}/d_{3,3,6,6})^2=14.1$. Here,
\begin{equation}
d_{p,l,p',l'}(z):=d_\text{eff}\mathcal{C}_{p,l}(z)c'_{p',l'}
\end{equation}
is the mode-dependent effective nonlinearity if only one of the harmonic modes is considered. Hence, the minimum possible ratio of the conversion efficiencies of the $\text{LG}_{0,0}$ and $\text{LG}_{3,3}$ mode is 7, assuming identical focusing and pump power as well as no pump-depletion. The maximum possible ratio is 14. Evidently, there is not a fixed ratio of the mode-dependent effective nonlinearities of two pump modes that would be measured in an experiment, here equal to the square-root of the ratio of the conversion efficiencies. It rather depends on the focusing because this determines the mode composition of the final harmonic field where each harmonic mode contributes with a different $d_{p,l,p',l'}$ value.

\section{Reconversion}\label{sec:reconversion}
For both pump modes, $\text{LG}_{0,0}$ as well as $\text{LG}_{3,3}$, harmonic power is occasionally reconverted into the fundamental field. In the case of the $\text{LG}_{0,0}$ mode, this is when the harmonic power briefly decreases for $\xi=50$. In the case of the $\text{LG}_{3,3}$ mode, the modes $\text{LG}_{0,6}$, $\text{LG}_{2,6}$ and $\text{LG}_{4,6}$ experience reconversion in almost each simulated case. In this section, we qualitatively investigate into which fundamental modes the harmonic modes are reconverted by using the PHE for the change in the fundamental field.

This PHE shows the same structure as Eq.\ \ref{DEQ:initialForm} and can also be rewritten similiar to Eq.\ \ref{DEQ:discreteHelmholtz} such that the fundamental field at the subsequent position $z+\Delta z$ is given by the interference of the fundamental field which propagated from $z$ to $z+\Delta z$ and the fundamental field which is emitted from the crystal polarisation at $z+\Delta z$. In the case of efficient conversion, this interference is destructive and results in the depletion of the pump field. For the qualitative analysis of the reconversion, only the mode-dependent part of the corresponding crystal polarisation $C^f$ is required. It is given by the following mixing of fundamental and harmonic field \cite{Boyd-nonlinearOptics}:
\begin{equation}
C^f(z)\propto A_2(r,\phi,z)A^*_1(r,\phi,z)\ \text{.}
\label{DEQ:reconversionInitialForm}
\end{equation}  
We assume $A_1$ to only contain the used pump mode while $A_2$ is a superposition of all excited harmonic modes. Hence, the crystal polarisation reads
\begin{equation}
C^f(z)\propto\left[\sum_{m=0}^{p}a_{2m,l'}(z)A^h_{2m,l'}(r,\phi,z)\right]A^*_{p,l}(r,\phi,z)\ \text{.}
\end{equation}
With
\begin{equation}
C^f_{p,l,p',l'}(z):=A^h_{p',l'}(r,\phi,z)A^*_{p,l}(r,\phi,z)\ \text{,}
\end{equation}
the $z$-independent overlap integral
\begin{align}
\begin{split}
\sigma_{p,l,p',l',p'',l''}&:=N_{p,l,p',l'}\ \Big|\!\int r A^*_{p'',l''}(r,\phi,z)C^f_{p,l,p',l'}(r,\phi,z)\ drd\phi\ \Big|^2
\end{split}
\end{align}
corresponds to the fraction of the harmonic mode $\text{LG}_{p',l'}$ which is reconverted into the fundamental mode $\text{LG}_{p'',l''}$ if the fundamental field is in the $\text{LG}_{p,l}$ mode. $N_{p,l,p',l'}$ is a normalisation factor. Tab.\ \ref{tab:overlapsForReconversion} shows that none of the harmonic modes completely reconverts into the original pump mode for either of the here analysed pump modes. Each harmonic mode rather reconverts into an individual superposition of fundamental modes which will act as additional pump modes that, in turn, excite individual superpositions of harmonic modes. In the regime of no pump-depletion and in cavity-enhanced SHG when the additional modes show no power buildup, this should, however, be a negligible effect.
\begin{table}[h!]
	\centering
	\caption{\bf Overlap integrals for reconversion}
	\begin{tabular}{cc}
		\hline\vspace*{-0.6cm}&\\\vspace*{0.1cm}
		$p,l$ , $p',l'$ , $p'',l''$ &\quad $\sigma_{p,l,p',l',p'',l''}$ \\
		\hline
		0,0 , 0,0 , 0,0 &\quad 0.7500 \\
		3,3 , 0,6 , 3,3 &\quad 0.3141 \\
		3,3 , 2,6 , 3,3 &\quad 0.1518 \\
		3,3 , 4,6 , 3,3 &\quad 0.1907 \\
		3,3 , 6,6 , 3,3 &\quad 0.5732 \\
		\hline
	\end{tabular}
	\label{tab:overlapsForReconversion}
\end{table}

\section{Double-pass and cavity-enhanced SHG}\label{sec:doublePass}
\begin{figure}[htbp]
	\centering
	\begin{tabular}{c|c}
		\captionsetup[subfigure]{oneside,margin={0cm,0cm}}
		\subfloat{\hspace*{-0.3cm}\includegraphics[trim=1cm 0cm 3cm 2cm,clip,width=0.5\linewidth]{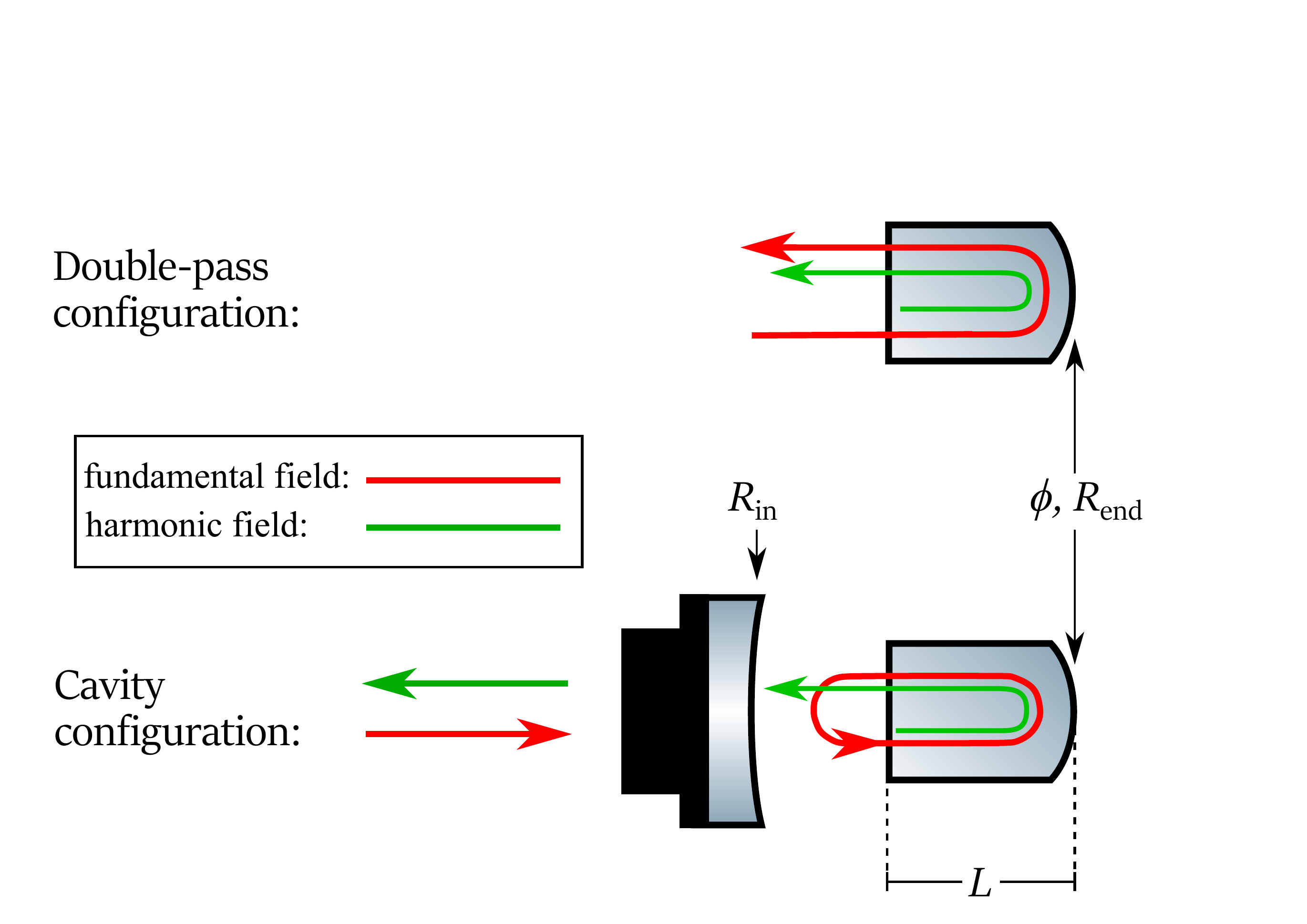}}&
		\captionsetup[subfigure]{oneside,margin={0cm,0cm}}
		\subfloat{\hspace*{-0.15cm}\includegraphics[trim=3cm 9.35cm 3cm 9.45cm,clip,width=0.48\linewidth]{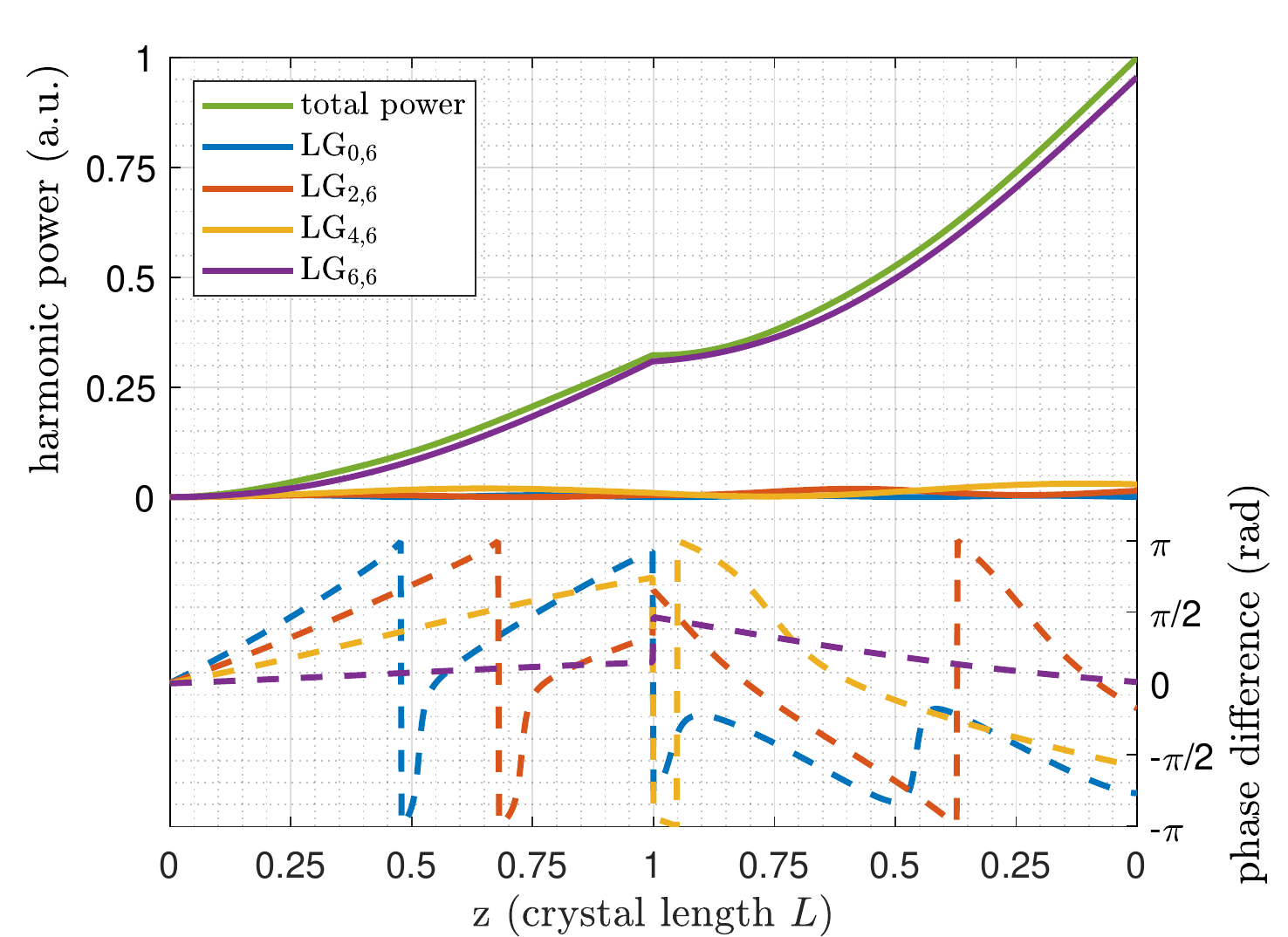}}
	\end{tabular}
	\caption{Left: Examplary schematic of the double-pass and cavity configuration. In both cases, the crystal end face acts as the (cavity end) mirror. Right: Numerical simulation of a double-pass SHG pumped by the $\text{LG}_{3,3}$ mode for $\xi=0.5$ and $\Delta n=0$. The waist is located at the centre of the crystal. The solid lines refer to the left axis and show the accumulated harmonic power. The dashed lines refer to the right axis and illustrate the phase difference between the harmonic field and the crystal polarisation for the harmonic $\text{LG}_{0,0}$ mode.}
	\label{fig:LG33doublePass}
\end{figure}

In gravitational wave detectors, the squeezed vacuum states in the $\text{LG}_{0,0}$ mode are generated in an optical parametric amplifier (OPA) via parametric down-conversion which is pumped by the harmonic output beam of an SHG. Below the OPA threshold, the generated squeezing level increases with the harmonic pump power. Hence, the SHG conversion efficiency is usually enhanced by the power buildup of the fundamental pump field in an optical linear cavity \cite{AdVirgoSqueezing} to achieve a high harmonic output power which allows for a high squeezing level. Such a cavity, used e.g.\ in \cite{CoherentControlBroadbandVacuumSqueezing} (see Fig.\ \ref{fig:LG33doublePass}), typically features an end mirror which is highly reflective at both the fundamental and harmonic frequency, and is designed such that the harmonic field is not resonating and coupled out in reflection (see below for examplary reflectivities). Thus, the harmonic field is generated in a double-pass through the crystal. The main two differences to a single pass as simulated above are that the fundamental and harmonic field usually experience different phase shifts under the reflection off the end mirror and that the second pass can be interpreted as a single pass with both non-zero initial harmonic power and non-zero initial phase differences. Fig.\ \ref{fig:LG33doublePass} shows the double-pass simulation for the $\text{LG}_{3,3}$ mode with no wavevector mismatch. Here, we assume the typical focusing regime of such an SHG cavity, $\xi=0.5$, and include the theoretical phase shift values for internal reflection of a coated \SI{7}{\percent} doped $\text{MgO:LiNbO}_3$ crystal as indicated by Laseroptik GmbH: $\phi_{\SI{1064}{nm}}=\SI{29.1}{\degree}$ and $\phi_{\SI{532}{nm}}=\SI{1.3}{\degree}$. Even though conversion efficiencies close to 1 can be achieved in cavity-enhanced SHG with respect to the fundamental input power, this simulation assumes that the regime of no pump-depletion is still a good approximation for the fundamental intra-cavity power. With a minimum power buildup factor of about 40 for the below simulated cavity (corresponding to the maximum achievable conversion efficiency), high conversion effciencies only correspond to a depletion of the intra-cavity pump power by about \SI{2.5}{\percent} or less during one roundtrip (double-pass). 

When the double-pass SHG is pumped by the $\text{LG}_{0,0}$ mode, the harmonic power and phase difference of the harmonic $\text{LG}_{0,0}$ mode will again qualitatively evolve identically to the harmonic $\text{LG}_{6,6}$ mode due to the same $\Delta\gamma_{p,l,p',l'}$ values.

According to the theoretical coating phase shifts, the reflection off the end mirror results in an increase of the phase difference by $2\phi_{\SI{1064}{nm}}-\phi_{\SI{532}{nm}}$. If the phase difference of any harmonic mode is close to $\pi$ or negative before the reflection, this increase can improve on the conversion around the point of reflection as it reduces the absolute value of the phase difference (see e.g. $\text{LG}_{0,6}$ in Fig.\ \ref{fig:LG33doublePass}). However, especially for the dominating modes which exhibit significantly smaller phase differences, this increase will deteriorate the conversion around the point of reflection. After the reflection, the phase differences evolve inverted due to the inverted propagation direction. Furthermore, the second pass generates more harmonic power because there is already harmonic power present at the start of this pass (see Eq.\ \ref{eqn:changeInPower}).

The simulation suggests that the $\text{LG}_{6,6}$ mode dominates the final harmonic field for the double-pass SHG in the $\xi\sim 0.5$ regime and we thus conclude the same for the output field of an SHG cavity as described above. The exact $\text{LG}_{6,6}$ contribution to the final harmonic field depends on the crystal temperature, which affects the wavevector mismatch as well as the phase shifts under reflection and is typically adjusted to optimise the conversion efficiency. The ratio of the pump mode-dependent effective nonlinearities is close to the maximum of 3.77 and, in turn, depends on the exact $\text{LG}_{6,6}$ contribution. Furthermore, as for the single-pass, the results for $\xi\sim 3$ would be similar.

\begin{figure}[htbp]
	\centering
	\hspace*{-0.4cm}\includegraphics[trim=3.1cm 11.5cm 4.2cm 11.7cm,clip,width=8cm]{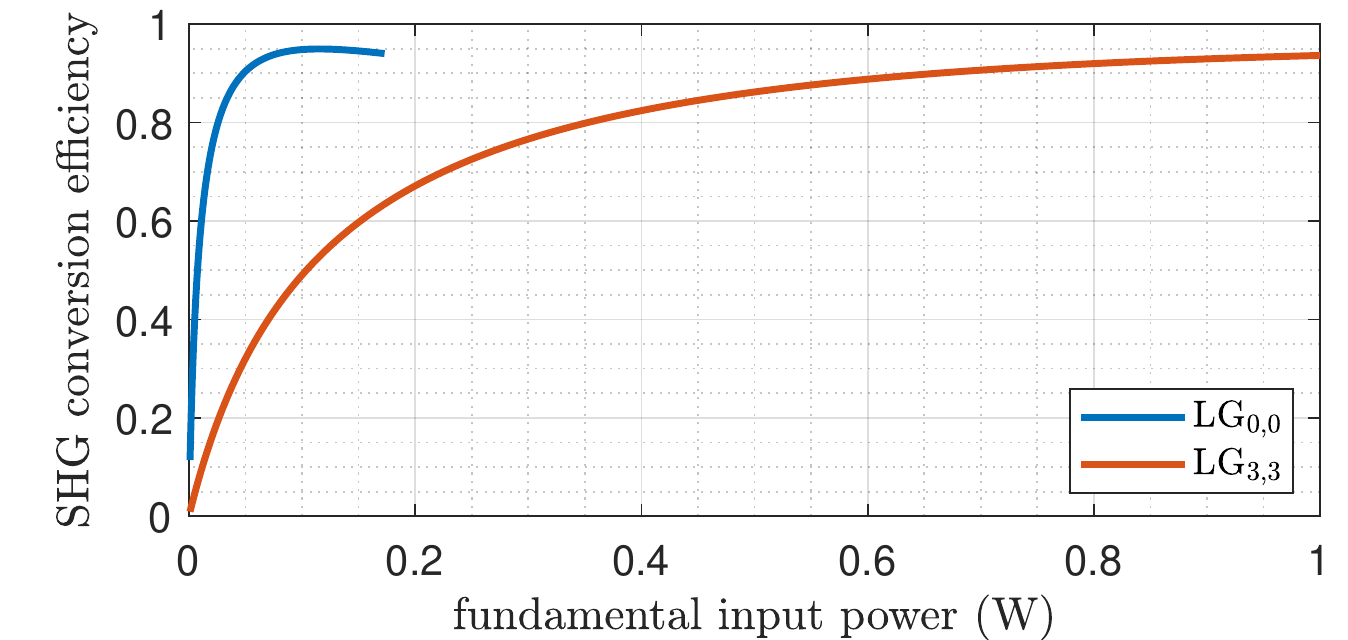}
	\caption{Numerical simulation of a cavity-enhanced SHG pumped by the $\text{LG}_{0,0}$ and $\text{LG}_{3,3}$ mode for $\xi=0.5$, $\Delta n=0$ and $d_{\text{eff},0,0}/d_{\text{eff},3,3}=3.7$.}
	\label{fig:SHGcavity}
\end{figure}

The Non-Linear Cavity Simulator (NLCS) from \cite{LastzkaPhD} calculates the conversion efficiency of an SHG cavity depending on the fundamental input power by using the following parameters: reflectivities of the cavity mirrors, crystal length, refractive index of the nonlinear medium, pump waist size and position with respect to the crystal centre, fundamental wavelength, absorption coefficients, wavevector mismatch and the effective nonlinearity. Fig.\ \ref{fig:SHGcavity} shows the comparison of the $\text{LG}_{0,0}$ and $\text{LG}_{3,3}$ mode in a cavity-enhanced SHG simulated with the NLCS for the following values: $\lambda_1=\SI{1064}{nm}$, $n=2.23$ (nominal refractive index for \SI{7}{\percent} doped $\text{MgO:LiNbO}_3$), $\Delta n=0$, $R_\text{end}=0.999$ (highly reflective end mirror at both involved frequencies), $R_{\text{in},\SI{1064}{nm}}=0.98$ (high reflectivity of incoupling mirror at fundamental frequency for power buildup), $R_{\text{in},\SI{532}{nm}}=0$ (harmonic field not resonating and coupled out in reflection), $L=\SI{6.5}{mm}$ (geometric crystal length), $w_0=\SI{31.4}{\micro\meter}$ (pump waist size in radius corresponding to $\xi=0.5$), $d_{\text{eff},0,0}=\SI{3}{\pico\meter\per\volt}$ (typical $d_\text{eff}$ for \SI{7}{\percent} doped $\text{MgO:LiNbO}_3$), $d_{\text{eff},3,3}=d_{\text{eff},0,0}/3.7$ (assuming that the harmonic $\text{LG}_{6,6}$ mode is almost exclusively present in the harmonic output field of the $\text{LG}_{3,3}$ SHG). No absorption loss is assumed. 

As an important result from our simulation, the ratio $d_{\text{eff},0,0}/d_{\text{eff},3,3}=3.7$ implies that the $\text{LG}_{3,3}$ SHG requires \num{13.69\pm 0.15} ($=3.7^2$) times the input power to achieve the same conversion efficiency as the $\text{LG}_{0,0}$ SHG. This factor was independent on the absolute $d_{\text{eff},0,0}$ value in the simulation and also remained the same when we assumed equal absorption losses in both cases.

\section{Conclusion}
We have first theoretically investigated the continuous-wave second harmonic generation pumped by an undepleted pump mode which generally excites a superposition of harmonic modes and secondly compared the $\text{LG}_{0,0}$ and $\text{LG}_{3,3}$ SHG in the single-, double-pass and cavity-enhanced configuration via numerical simulations. We have derived a general equation which explains the fundamentally different conversion efficiencies of two pump modes as an effect of their intensity distributions. The same equation shows how the Gouy phase causes the harmonic modes which are excited by a single pump mode to exhibit different phase matching conditions. These influences have been illustrated by the simulations for different focusing regimes with a special focus on the phase relations between the excited harmonic modes and the crystal polarisation.

Our comparison predicts that the $\text{LG}_{3,3}$ mode requires about 14 times the fundamental input power compared to the $\text{LG}_{0,0}$ mode to achieve the same conversion efficiency in an ideal cavity-enhanced SHG in the range from the typical to the optimum focusing regime of $\xi\sim 0.5$ and $\xi\sim 3$, respectively. This corresponds to a ratio of the pump mode-dependent effective nonlinearities of about 3.7. Furthermore, we have shown that an $\text{LG}_{3,3}$ SHG in this focusing range will mainly generate the harmonic $\text{LG}_{6,6}$ mode ($>\SI{90}{\percent}$).

Since squeezed vacuum states of light for the gravitational wave detectors are generated using parametric down-conversion, which can be seen as a reversed SHG, our results also suggest that only the $\text{LG}_{6,6}$ mode can efficiently pump the optical parametric amplifier to produce squeezed vacuum states in the $\text{LG}_{3,3}$ mode in the range from $\xi\sim 0.5$ to $\xi\sim 3$. We expect this process to require about 14 times the pump power to generate the same squeezing level in the $\text{LG}_{3,3}$ mode, when pumped by a pure harmonic $\text{LG}_{6,6}$ mode, compared to the $\text{LG}_{0,0}$ mode, when pumped by a pure harmonic $\text{LG}_{0,0}$ mode. In a cavity configuration as described above, there is, however, also the option to increase the Finesse for the fundamental squeezed field via a higher reflectivity of the incoupling mirror. This is not an independent parameter because an increasing incoupling reflectivity will eventually significantly reduce the achievable squeezing level via a reduced espace efficiency. However, the Finesse should not have to be increased up to this regime such that $\text{LG}_{0,0}$ squeezing levels are theoretically possible in the $\text{LG}_{3,3}$ mode for the same pump power. Since the increased Finesse only applies to the squeezed vacuum field, there will be no significant increase in the circulating fundamental power and no associated increased risk of photo-thermal damage in the nonlinear crystal.

\section*{Disclosures}

The authors declare no conflicts of interest.

\section*{Funding}

Funded by the Deutsche Forschungsgemeinschaft (DFG, German Research Foundation) under Germany’s Excellence Strategy – EXC-2123 QuantumFrontiers – 390837967.

\bibliography{SHGwithLG33}

\begin{thebibliography}{10}
\newcommand{\enquote}[1]{``#1''}

\bibitem{Intro:aLIGO}
{The LIGO Scientific Collaboration}, \enquote{{A}dvanced {LIGO},}
  {\protect\JournalTitle{Classical and Quantum Gravity}} \textbf{32} (2015).

\bibitem{Intro:AdV}
{The Virgo Collaboration}, \enquote{{A}dvanced {V}irgo: a second-generation
  interferometric gravitational wave detector,}
  {\protect\JournalTitle{Classical and Quantum Gravity}} \textbf{32} (2015).

\bibitem{Intro:reduceThermalNoise}
J.-Y. Vinet, \enquote{Reducing thermal effects in mirrors of advanced
  gravitational wave interferometric detectors,}
  {\protect\JournalTitle{Classical and Quantum Gravity}} \textbf{24} (2007).

\bibitem{Birmingham-highpurity}
P.~Fulda, K.~Kokeyama, S.~Chelkowski, and A.~Freise, \enquote{Experimental
  demonstration of higher-order {L}aguerre-{G}auss mode interferometry,}
  {\protect\JournalTitle{Physical Review D}} \textbf{82} (2010).

\bibitem{prospects}
S.~Chelkowski, S.~Hild, and A.~Freise, \enquote{Prospects of higher-order
  laguerre-gauss modes in future gravitational wave detectors,}
  {\protect\JournalTitle{Physical Review D}} \textbf{79}, 122002 (2009).

\bibitem{highPowerHighPurity}
L.~Carbone, C.~Bogan, P.~Fulda, A.~Freise, and B.~Willke, \enquote{Generation
  of {H}igh-{P}urity {H}igher-{O}rder {L}aguerre-{G}auss {B}eams at {H}igh
  {L}aser {P}ower,} {\protect\JournalTitle{Physical Review Letters}}
  \textbf{110} (2013).

\bibitem{Noack}
A.~Noack, C.~Bogan, and B.~Willke, \enquote{Higher-order {L}aguerre-{G}auss
  modes in (non-) planar four-mirror cavities for future gravitational wave
  detectors,} {\protect\JournalTitle{Optics Letters}} \textbf{42}, 751--754
  (2017).

\bibitem{MichelsonInterferometer_LG33}
A.~Gatto, M.~Tacca, F.~K\'ef\'elian, C.~Buy, and M.~Barsuglia,
  \enquote{{F}abry-{P}\'erot-michelson interferometre using higher-order
  {L}aguerre-{G}auss modes,} {\protect\JournalTitle{Physical Review D}}
  \textbf{90} (2014).

\bibitem{10mHighFinesseCavity_LG33}
B.~Sorazu, P.~J. Fulda, B.~W. Barr, A.~S. Bell, C.~Bond, L.~Carbone, A.~Freise,
  S.~Hild, S.~H. Huttner, J.~Macarthur, and K.~A. Strain,
  \enquote{{E}xperimental test of higher-order {L}aguerre-{G}auss modes in the
  10m glasgow prototype interferometer,} {\protect\JournalTitle{Classical and
  Quantum Gravity}} \textbf{30} (2013).

\bibitem{degeneracyHighFinesseCavity}
C.~Bond, P.~Fulda, L.~Carbone, K.~Kokeyama, and A.~Freise, \enquote{{H}igher
  order {L}aguerre-{G}auss mode degeneracy in realistic, high finesse
  cavities,} {\protect\JournalTitle{Physical Review D}} \textbf{84} (2011).

\bibitem{AdVirgoSqueezing}
{The Virgo Collaboration}, \enquote{{I}ncreasing the {A}strophysical {R}each of
  the {A}dvanced {V}irgo {D}etector via the {A}pplication of {S}queezed
  {V}acuum {S}tates of {L}ight,} {\protect\JournalTitle{Physical Review
  Letters}} \textbf{123} (2019).

\bibitem{BoydKleinmann}
G.~Boyd and D.~A. Kleinmann, \enquote{{P}arametric {I}nteraction of {F}ocused
  {G}aussian {L}ight {B}eams,} {\protect\JournalTitle{Journal of Applied
  Physics}} \textbf{39}, 3597--3639 (1968).

\bibitem{lastzkaGouy}
N.~Lastzka and R.~Schnabel, \enquote{The {G}ouy phase shift in nonlinear
  interactions of waves,} {\protect\JournalTitle{Optics Express}} \textbf{15},
  7211--7217 (2007).

\bibitem{SHGmodeDiscrimination}
V.~Delaubert, M.~Lassen, D.~R.~N. Pulford, H.-A. Bachor, and C.~C. Harb,
  \enquote{{S}patial mode discrimination using second harmonic generation,}
  {\protect\JournalTitle{Optics Express}} \textbf{15}, 5815--5826 (2007).

\bibitem{SHGindividualPhaseMatching}
P.~Buchhave and P.~Tidemand-Lichtenberg, \enquote{{G}eneration of higher order
  {G}auss-{L}aguerre modes in single-pass 2nd harmonic generation,}
  {\protect\JournalTitle{Optics Express}} \textbf{16}, 17952--17961 (2008).

\bibitem{CoherentControlBroadbandVacuumSqueezing}
S.~Chelkowski, H.~Vahlbruch, K.~Danzmann, and R.~Schnabel, \enquote{Coherent
  control of broadband vacuum squeezing,} {\protect\JournalTitle{Physical
  Review A}} \textbf{75} (2007).

\bibitem{Siegmann-Laser}
A.~E. Siegmann, \emph{Lasers} (University Science Books, 1986).

\bibitem{Boyd-nonlinearOptics}
R.~W. Boyd, \emph{Nonlinear Optics} (Academic Press, 2008).

\bibitem{SHG1997}
J.~Courtial, K.~Dholakia, L.~Allen, and M.~J. Padgett, \enquote{Second-harmonic
  generation and the conversion of orbital angular momentum with high-order
  {L}aguerre-{G}auss modes,} {\protect\JournalTitle{Physical Review A}}
  \textbf{56}, 4193--4196 (1997).

\bibitem{LastzkaPhD}
N.~Lastzka, \enquote{Numerical modelling of classical and quantum effects in
  non-linear optical sysmtes,} Ph.D. thesis, Leibniz Universit\"at Hannover
  (2010).

\end{thebibliography}






\end{document}